\documentclass[11pt]{article}
\pdfoutput=1

\usepackage{jcapmod}
\usepackage[usenames]{xcolor}
\usepackage[english]{babel}
\usepackage[small]{caption}
\usepackage{verbatim}
\usepackage{multirow}
\usepackage{relsize}
\usepackage{amsmath,amssymb,amsbsy,amstext,amsfonts}
\usepackage{epsfig,multicol,bbm,simplewick}
\usepackage{mathtools}
\usepackage{enumitem}
\usepackage{csquotes}
\usepackage{tensor}
\usepackage{fix-cm}
\usepackage{booktabs}
\usepackage{graphicx}
\usepackage{bm}
\usepackage{subfigure}
\usepackage{float}
\usepackage[explicit]{titlesec}
\usepackage{enumitem}
\usepackage{setspace}
\usepackage{blindtext}
\usepackage{textpos}
\usepackage{etoolbox}
\usepackage[linktocpage=true]{hyperref}
\usepackage{anyfontsize}
\usepackage{listings,xfp}
\usepackage[dotinlabels]{titletoc}
\usepackage[normalem]{ulem}
\usepackage{esvect}
\DeclareMathAlphabet\mathbfcal{OMS}{cmsy}{b}{n}

\dottedcontents{section}[1.2em]{}{1.2em}{6pt}
\dottedcontents{subsection}[5em]{}{2em}{6pt}

\hypersetup{colorlinks=true,linkcolor=blue,citecolor=blue,urlcolor=blue}
\usepackage[a4paper, total={6.5in, 9.5in}]{geometry}
\renewcommand{\thesection}{\arabic{section}}
\definecolor{airforceblue}{rgb}{0.36, 0.48, 0.84}
\definecolor{purplemath}{rgb}{0.5, 0, 0.5}

\makeatletter
\def\@xfootnote[#1]{%
  \protected@xdef\@thefnmark{#1}%
  \@footnotemark\@footnotetext}
\makeatother

\newcommand{\be}{\begin{equation}}
\newcommand{\ee}{\end{equation}}
\newcommand{\bea}{\begin{eqnarray}}
\newcommand{\eea}{\end{eqnarray}}

\newcommand{\kc}{k_\sigma}

\newcommand*\diff{\mathop{}\!\mathrm{d}}

\newcommand{\eq}[1]{(\ref{#1})}

\begin{document}

\begin{titlepage}

\setcounter{page}{1} \baselineskip=15.5pt \thispagestyle{empty}

\noindent IFT-UAM/CSIC-24-75 \hfill DESY-24-071

\vspace{0.1cm}

\bigskip\
\begin{center}
{\fontsize{18}{30}\selectfont \bf Non-Gaussian tails without stochastic inflation} 
\end{center}

\begin{NoHyper}
\begin{center}
{\fontsize{13}{30}\selectfont
Guillermo Ballesteros,$^{1,2}$\footnote[]{guillermo.ballesteros@uam.es}
Thomas Konstandin,$^3$\footnote[]{thomas.konstandin@desy.de}
\\ \vspace{0.05in}
Alejandro P\'erez Rodr\'iguez,$^{1,2}$\footnote[]{alejandro.perezrodriguez@uam.es}
Mathias Pierre,$^3$\footnote[]{mathias.pierre@desy.de}
and Juli\'an Rey$^3$\footnote[]{julian.rey@desy.de}}
\end{center}
\end{NoHyper}

\vspace{0.1cm}

\begin{center}
{\small
$^1$\textsl{Departamento de F\'isica Te\'orica, Universidad Aut\'onoma de Madrid (UAM),\\
Campus de Cantoblanco, 28049 Madrid, Spain}\\ \vspace{0.14cm}
$^2$\textsl{Instituto de F\'isica Te\'orica (IFT) UAM-CSIC, Campus de Cantoblanco, 28049 Madrid, Spain}\\
$^3$\textsl{Deutsches Elektronen-Synchrotron DESY, Notkestr. 85, 22607 Hamburg, Germany}
}
\end{center}

\vspace{0.6cm}

\begin{center}
{\large \bf Abstract}
\end{center}

\noindent 
We show, both analytically and 
numerically, 
that 
non-Gaussian tails in the probability density function 
of curvature perturbations arise in ultra-slow-roll inflation
from the $\delta N$ formalism, without invoking
stochastic inflation. 
Previously reported discrepancies between both approaches are a consequence of not correctly accounting for momentum perturbations. Once they are taken into account, both approaches agree to an excellent degree. 
The shape of the tail 
depends strongly on the phase space of inflation.
\vspace{0.4cm}

\tableofcontents

\end{titlepage}

\section{Introduction}

There is a broad consensus that current observational bounds allow primordial black holes (PBHs) with masses between $10^{-16}\, M_{\odot}$ and $10^{-12}\, M_{\odot}$ to account for all the dark matter of the Universe \cite{Green:2020jor}. The most popular idea for the formation of PBHs that could constitute the dark matter assumes the existence of very large density fluctuations in the early Universe (before primordial nucleosynthesis), which collapse under their own gravitational pull. Indeed, if the collapse is assumed to start during radiation domination, each fluctuation needs to overcome a threshold for the relative overdensity of nearly $\mathcal{O}(1)$ 
in order to become a black hole. The statistical properties of these fluctuations are the main factor determining the current PBH abundance and mass distribution.
Thinking in terms of Fourier modes to describe the early Universe inhomogeneities, the collapse towards a black hole starts to occur as soon as the wavelength of a fluctuation above the threshold is comparable to the Hubble radius. This wavelength is the characteristic scale that determines the mass of the resulting black holes.

The possible origin of these large density fluctuations is uncertain. A popular way to realize the idea is to play with the dynamics of inflation in such a way that curvature fluctuations acquire a large variance at adequate distance scales. Using linear cosmological perturbation theory to relate density and curvature fluctuations and assuming Gaussianity of the latter, a simple estimate (see e.g.\ \cite{Ballesteros:2017fsr}) shows that a power spectrum of curvature perturbations $\sim 10^{-2}$ --i.e.\ about seven orders of magnitude larger than fluctuations at the scales we are able to probe with the Cosmic Microwave Background (CMB)-- is needed to account for all dark matter. One way to attain this consists in introducing a period of ultra-slow roll inflation (USR) \cite{Kinney:2005vj}, during which Fourier modes of curvature fluctuations experience super-Hubble growth~\cite{Ivanov:1994pa}. Concrete models that implement this mechanism are e.g.\,the ones proposed in \cite{Ballesteros:2017fsr} and~\cite{Ballesteros:2020qam}. 

It is known that in USR inflation, Fourier modes with small amplitude acquire some amount of local non-Gaussianity \cite{Taoso:2021uvl}. This already indicates that the previous Gaussian estimate for the abundance of PBHs in such models is too simplistic. In addition, several recent studies have highlighted the appearance in models of USR of non-Gaussian (exponential) tails in the probability distribution function (PDF) of $\zeta$ for large values~\cite{Ezquiaga:2019ftu,Figueroa:2021zah,Pi:2022ysn,Domenech:2023dxx}. These results have been obtained using mainly two different (but related) frameworks that go beyond standard cosmological perturbation theory and several computational techniques: the $\delta N$ {\it formalism} and the {\it stochastic} $\delta N$ {\it formalism}.\footnote{In what follows we will often use the term {\it classical} $\delta N$ formalism to refer to the non-stochastic version of it. The term classical is used here in contrast to stochastic, and in no way implies that the fluctuations in the $\delta N$ formalism are not quantum in nature.} In this work we aim to shed further light on the appearance of these tails. Specifically, we are interested in understanding under which conditions (if any) the results obtained with these two frameworks differ.

In essence, the $\delta N$ formalism exploits a non-linear relation \cite{Atal:2019cdz} between the inflaton fluctuations, $\delta \phi$, and the curvature perturbation on uniform density slices $\zeta$.\footnote{See Appendix \ref{ap:deltan} and ref.\ \cite{Maldacena:2002vr} for the definition of $\zeta$.} Once this non-linear relation is established and {\it if} the PDF of $\delta\phi$ is known, the PDF of $\zeta$ can be easily obtained. If $\delta\phi$ is {\it assumed} to be a Gaussian variable, its non-linear relation to $\zeta$ makes the PDF of the latter non-Gaussian. In particular, in the case of USR, an exponentially decaying tail arises for large $\zeta$.
Therefore, within this formalism, the appearance of a non-Gaussian tail in the PDF of $\zeta$
is just the consequence of a change of variables. The stochastic $\delta N$ formalism adds another twist to this argument by incorporating the framework of {\it stochastic inflation} \cite{Starobinsky:1986fx}. This framework is used to study the effect of short-wavelength modes on the dynamics of long-wavelength modes. This backreaction is included via stochastic noise terms acting on the dynamical equations of the latter. Within this framework (whose full non-linear implementation is highly costly in computational terms \cite{Figueroa:2021zah}), the appearance of exponential tails has been presented as a consequence of {\it quantum diffusion}, see  e.g.~\cite{Ezquiaga:2019ftu}. A natural question to ask is how different are the tails obtained in these two formalisms \cite{Tomberg:2023kli}.\footnote{Stochastic inflation has been shown to give identical results to those of perturbation theory at the linear level, see e.g.\,\cite{Cruces:2018cvq,Ballesteros:2020sre,Cruces:2021iwq}.}

In our work we use two different implementations (models) of USR to 
carefully analyze the relevance of the phase space of
inflation in order to understand the practical differences between the two formalisms. In the process of doing so, we discuss the validity and convenience of using a specific approximation (the {\it unperturbed-trajectory approximation}) that simplifies the treatment of the fluctuations of the inflaton assuming they follow, at all times, 
the unperturbed solution of the background equations of motion. 

In the next section we introduce our notation and the two models we use. One of them ({\it Model~1}) is a piecewise implementation of the USR phase and the other ({\it Model 2}),
first presented in \cite{Ballesteros:2017fsr},
involves a polynomial potential for the inflaton. In Section \ref{sec:deltan} we discuss the application of the classical $\delta N$ formalism to the calculation of the PDF of $\zeta$ assuming Gaussian inflaton fluctuations, first by using the unperturbed-trajectory approximation and then by performing the full calculation in both models. We find that the unperturbed-trajectory approximation works well only for {\it Model 2} and it fails for {\it Model 1} due to the more complex shape of its phase space. In Section \ref{sec:stochastic} we discuss the stochastic $\delta N$ formalism and compute the PDF of $\zeta$ for both models. An analogous approximation to the one we explored for the classical $\delta N$ formalism can also be used now, but this time it relates the stochastic terms.
Again, we will find a qualitative difference among the behaviours of both models, which can be understood in terms of a phase space analysis. We present our conclusions in Section~\ref{sec:conc}. An appendix is devoted to a more detailed presentation of the $\delta N$ formalism which includes several technical aspects that we do not discuss in the main body of the text. We use natural units and set the reduced Planck mass to 1. 

\section{Inflationary setup}
\label{sec:inflation}

Our inflationary setup consists of a single scalar field $\phi$ with an action given by
\begin{equation} \label{inflaset}
S = \int \diff^4x \sqrt{-g}
\bigg[\frac{1}{2}R+\frac{1}{2}(\partial_\mu\phi)^2-V(\phi)\bigg],
\end{equation}
where $g$ denotes the determinant of the spacetime metric, $R$ the Ricci scalar, and $V$ the
potential. We have also set the (reduced)  Planck mass to 1. We work with the following sign convention for the FLRW metric
\begin{equation}
\diff s^2=-\diff t^2+a(t)^2\diff {\bm x}^2,
\end{equation}
where $a(t)$ denotes the scale factor. {With this metric, the equations of motion for the (spatially homogeneous) field, $\phi$, and its (also homogeneous) conjugate momentum $\pi$} are
\begin{equation}
\label{eq:EoMphi}
\phi' = \pi
\qquad{\rm and}\qquad
\pi' = -(3-\epsilon)(\pi+\partial_{\phi}\log V) \, ,
\end{equation}
where primes denote derivatives with respect to the number of $e$-folds 
\begin{align}
\diff N=H\diff t \, ,
\end{align} 
with $H\equiv \dot a/a$ and we use the following definition for the slow-roll parameters,
\begin{equation}
\label{eq:sr-pars}
\epsilon = -\frac{H'}{H}=\frac{1}{2}\pi^2
\qquad{\rm and}\qquad
\eta=\epsilon-\frac{\epsilon'}{2\epsilon} = -\frac{\ddot{\phi}}{{\dot\phi} H}\,,
\end{equation}
where dots indicate derivatives with respect to cosmic time, $t$. In principle, \eqref{eq:EoMphi} is just a set of equations with no specified initial conditions. A specific solution to these equations is what we will call the \emph{unperturbed trajectory}, denoted by $\bar{\phi}$ and $\bar{\pi}$. By this, we refer to a solution of \eqref{eq:EoMphi} with initial conditions set in the far past (i.e.\ for $\phi$ larger than those corresponding to the USR part of the potential) and that has reached the SR attractor before entering the USR phase.\footnote{Since the SR part of the potential indeed corresponds to an attractor in phase space, there is a wide range of initial conditions that, if set sufficiently far in the past, satisfy this condition.} This corresponds to what is usually called the \emph{background solution} in standard perturbation theory, and it is depicted as a black dashed line in the phase spaces in Fig.\ \ref{fig:phase}. 

\begin{figure}
\centering
\includegraphics[width=0.49 \textwidth]{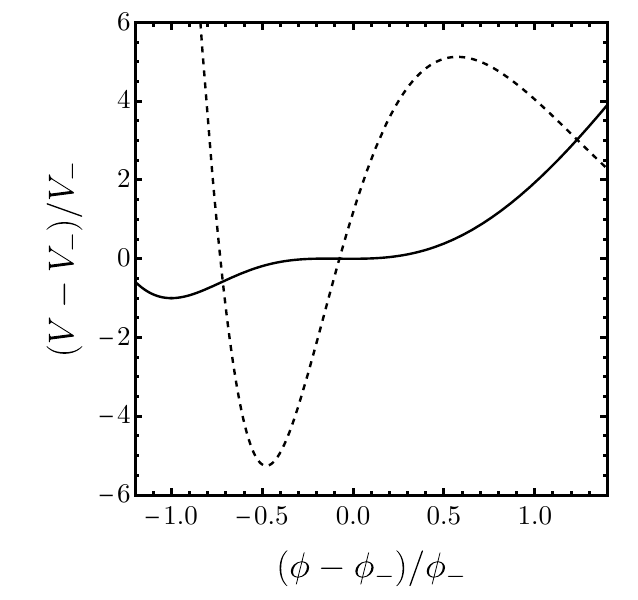} 
\includegraphics[width=0.49 \textwidth]{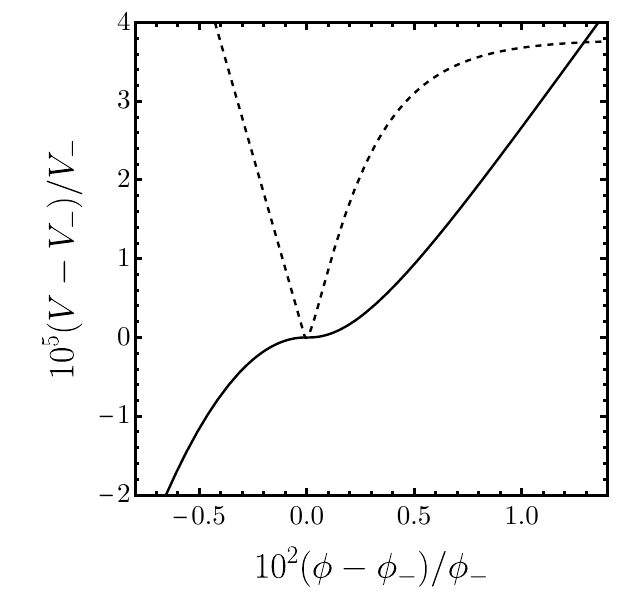} 
\caption{
\it {Potential (solid line) and second derivative $\diff^2V / \diff\phi^2$ (dashed line) for Model 1 (left panel) and Model 2 (right panel). $\phi_{-}$ denotes the location of the local minimum of the potential, and $V_{-}$ the corresponding value.}
}
\label{fig:potentials}
\end{figure}

Given explicit expressions for the slow-roll parameter $\eta$ as a function of time and an initial condition for $\epsilon$, the potential can be obtained as an integral over the number of $e$-folds as
\begin{equation}
V(N)=V(N_\star)\exp
\bigg[
2\int_{N_\star}^N\,\epsilon(\tilde N)\, \frac{\eta(\tilde N)-3}{3-\epsilon(\tilde N)}
\diff \tilde{N}
\bigg]\,.
\label{eq:pot_slowroll}
\end{equation}
In this expression $N_\star$ is just some reference $e$-fold value.

We are interested in models of inflation characterized by a potential, $V$, featuring a non-stationary ($\partial_\phi V\neq 0$) inflection
point 
so that the inflaton experiences an ultra-slow-roll (USR) phase, defined by $\eta \gtrsim 3/2$ and followed by a constant-roll (CR) phase, defined by $(\epsilon'/\epsilon)'\simeq 0$ and negative $\eta$.\footnote{USR is often defined in the literature as the condition $\ddot\phi+3H\dot\phi\simeq 0$. This corresponds to $\eta\simeq 3/2$ (and $\epsilon\ll 1$), which implies $V'\ll \ddot\phi$. In this work we adopt a broader definition, using the term USR to refer to any phase with $\eta\gtrsim 3/2$,
since for these values of $\eta$ the curvature perturbation grows exponentially, see Eq.\,(\ref{eq:second_mode}).}
Before reaching the inflection point, the inflaton undergoes a conventional slow-roll (SR) phase, where both $\epsilon$ and $\eta$ are small. As the field approaches the local maximum of the potential,
it decelerates, and $\eta$ becomes large.
{Indeed, approximating the potential in a neighbourhood of its maximum, $\phi_0$, by a parabola and assuming the field decreases as inflation proceeds, if we neglect $\epsilon$ in Eq.\ (\ref{eq:EoMphi}) and if $\phi'_0\equiv \phi'(N_0)\neq 0$ at $\phi_0$, then 
\begin{align}\label{linear_eom}
\phi\simeq\phi_0+ \frac{\phi'_0}{\sqrt{9-12\eta_V}}\left[-e^{-\eta_+(N-N_0)\,}+e^{-\eta_{-}(N-N_0)}\right]\,,
\end{align}
where $\eta_V=\partial_\phi^{\,2} V/V<0$ is a constant to a good degree of approximation and, in the limit $\phi\gg \phi_0$ we have that $\eta\simeq\eta_+\equiv\frac{3}{2}\left(1+\sqrt{1-4\eta_V/3}\right)$, whereas for $\phi\ll\phi_0$, $\eta\simeq\eta_{-}\equiv\frac{3}{2}\left(1-\sqrt{1-4\,\eta_V/3}\right)$. Therefore, the inflaton climbs up the potential in USR (with $\eta > 0$), and afterwards, it reaches a phase of CR (with $\eta < 0$). As noted in \cite{Karam:2022nym}, the values of $\eta$ in the two phases are related by the Wands duality \cite{Wands:1998yp}, such that their sum is $3$.
}

\begin{figure}
\centering
\includegraphics[width=0.49 \textwidth]{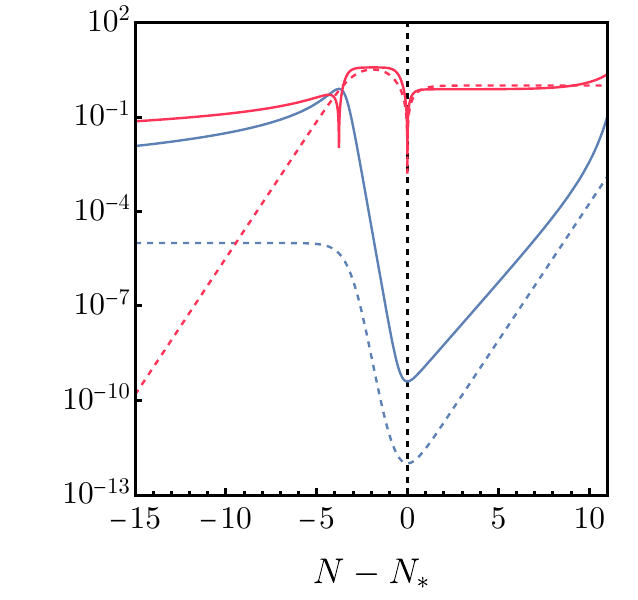} 
\includegraphics[width=0.49 \textwidth]{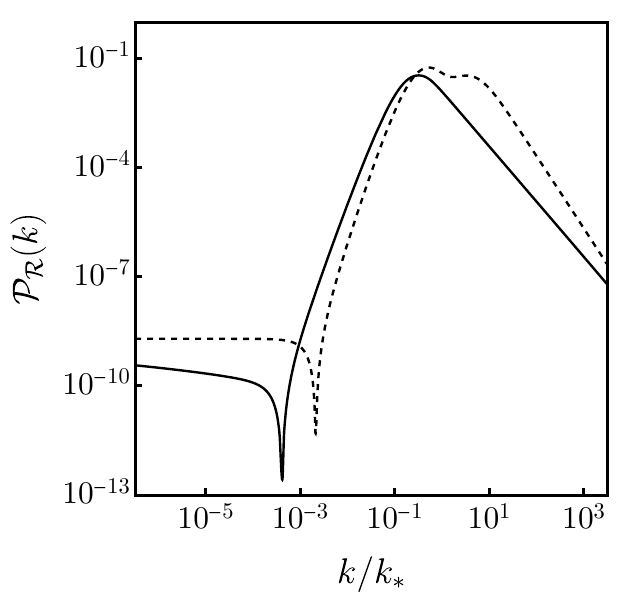} 
\caption{
\it Left panel: Slow-roll parameters $\epsilon$ (blue) and $|\eta|$ (red) for {\it Model 1} (dashed lines) and {\it Model 2} (solid lines). Right panel: Scalar power spectra corresponding to the two models in the left panel. $N_*$ is the time at which $\eta=3/2$ in the USR-CR transition along the unperturbed trajectory, see the comment below \eqref{eq:second_mode}.
}
\label{fig:model}
\end{figure}

In what follows, we will consider two different ways of implementing a phase of USR inflation. We refer to them as {\it Model 1} and {\it Model 2}.  The corresponding potentials are depicted in Fig.~\ref{fig:potentials}.  Their slow-roll parameters are shown in the left panel of Fig.\,\ref{fig:model}, and the corresponding linear power spectra of curvature fluctuations, in the right panel. 
\\

\noindent {$\star$ \bf Model 1.}\quad The first model consists of a phenomenological parameterization of $\eta$ as a function of the number of $e$-folds:\footnote{Similar parameterizations have been used before, see e.g.\ \cite{Taoso:2021uvl}. The difference between ours and the one in \cite{Taoso:2021uvl} is on the phase preceding USR. In our case $\eta$ gradually drops towards zero in the limit $N\rightarrow -\infty$, whereas the parameterization of \cite{Taoso:2021uvl} drops to zero more abruptly.} 
\begin{equation}
\eta(N) = \frac{\eta_{\rm USR}}{4}\Big[1+\tanh(N-{N}_{1})\Big]\Big[1+\tanh({N}_{2}-N)\Big]
+\frac{\eta_{\rm CR}}{2}\Big[1+\tanh{(N-{N}_{2})}\Big].
\label{eq:step_model}
\end{equation}
For our numerical example we choose $\eta_{\rm USR}=4$, $\eta_{\rm CR}=-1$ and ${N}_{2}-{N}_{1}=2.3$.
The USR phase (as defined above, with $\eta > 3/2$) sets in at
$N-N_{2}\simeq -2.54$
and ends at
$N\simeq N_2$.
In between these two instants, there is an interval (from 
$N-N_{2}\simeq -1.64$
to
$N-N_{2}\simeq -0.79$) during which $\eta$ is just above $3$ (and nearly constant). The subsequent 
CR phase begins approximately at $N\simeq {N}_{2}+3$.  In the limit $N\rightarrow -\infty$, $\eta$ tends rapidly to zero, as shown in Fig.\ \ref{fig:model}. The definition of $\eta$ in Eq.\ \eq{eq:sr-pars} can be integrated to obtain $\epsilon$, up to an initial condition which we choose in such a way that $\epsilon=10^{-5}$ in the far past. The potential can be obtained via Eq.\ (\ref{eq:pot_slowroll}), with $V(N_\star)$ chosen in such a way that the power spectrum is correctly normalized on CMB scales (again, in the far past). We take $\mathcal{P}_\mathcal{\zeta}(k_{\rm CMB})=2\times 10^{-9}$ at $k_{\rm CMB}=6.3\times 10^{-17}k_*$, where $k_*=a(N_*)H(N_*)$ and $N_*$ denotes the time at which $\eta\simeq 3/2$ after USR, see the comment below \eqref{eq:second_mode}.\footnote{Notice here that we do not impose the spectral index $n_s$ of the power spectrum at CMB scales to reproduce the one measured by the Planck collaboration since those scales are much larger than those relevant for PBH formation and we are interested only on the local dynamics around the inflection point of the potential.}
\\

\noindent {$\star$ \bf Model 2.}\quad The second model we study is the one presented in \cite{Ballesteros:2017fsr,Ballesteros:2020qam}, in which a scalar field $\chi$ with a fourth order polynomial potential couples to the Ricci scalar $R$ via a term $\xi \sqrt{-g} \chi^2 R$ in the Lagrangian. In the Einstein frame
\begin{equation}
U(\chi)
=
\frac{\lambda\chi^4}{4!(1+\xi\chi^2)^2}
\bigg[
3
+\xi^2\chi_0^4
-8(1+c_3)\frac{\chi_0}{\chi}
+2(1+c_2)(3+\xi\chi_0^2)\frac{\chi_0^2}{\chi^2}
\bigg],
\label{eq:poly_model}
\end{equation}
and $\chi$ has a non-canonical kinetic term. The coefficients of the quartic polynomial have been rewritten in a convenient way so that the potential features an inflection point at $\chi_0$ for $c_2=c_3=0$. The relation
\begin{equation}
\phi
=
\sqrt{\frac{1+6\xi}{\xi}}\sinh^{-1}
\Big[\chi\sqrt{\xi(1+6\xi)}\Big]
-\sqrt{6}\tanh^{-1}
\bigg[\frac{\sqrt{6}\xi\chi}{\sqrt{1+\xi(1+6\xi)\chi^2}}\bigg] \,,
\end{equation}
allows to work with the field $\phi$, which is canonically normalized.
We choose $c_2=0.011$, $c_3=0.0089$, $\chi_0=1$, $\xi=0.325479$ 
and $\lambda$ so that the power spectrum is correctly normalized on CMB scales \cite{Ballesteros:2020qam}. Again, we take $\mathcal{P}_\mathcal{\zeta}(k_{\rm CMB})=2.1\times 10^{-9}$ at $k_{\rm CMB}=6.3\times 10^{-17}k_*$.

\bigskip
Physical properties of the universe such as the CMB temperature fluctuations are related to inflation via the (linear) comoving curvature perturbation $\mathcal{R}$,\footnote{In flat gauge (see Appendix \ref{ap:deltan}), the {\it linear} comoving curvature perturbation reduces to $\mathcal{R}=-\delta\phi/\phi'$, where $\delta\phi$ is the inflaton fluctuation in that gauge.  At linear order in perturbation theory and for single-field inflation, $\mathcal{R}$ and $\zeta$ (which is the variable used in the bulk of this work) agree for frozen super-Hubble Fourier modes. } see e.g.\ \cite{Baumann:2022mni,Ballesteros:2018wlw}, which is gauge invariant at linear order and whose linearized Fourier modes obey the equation 
\begin{equation}
\mathcal{R}_k''+(3+\epsilon-2\eta)\mathcal{R}_k'+\frac{k^2}{a^2H^2}\mathcal{R}_k=0\,.
\label{eq:mukhanov-sasaki}
\end{equation}
In this equation, $H=\dot{a}/a$ and $k$ denotes the comoving wavenumber of the Fourier mode $\mathcal{R}_k$. At early times during inflation the modes are assumed to be in the Bunch-Davies vacuum:
\begin{equation}
\mathcal{R}_k=-\frac{1}{a\phi'\sqrt{2k}}\exp\bigg(-ik\int \frac{\diff N}{aH}\bigg).
\end{equation}
Let us assume that $\epsilon\ll 1$ (so $H'\simeq 0$) and $\eta$ is constant, which holds both during USR and CR. Then, for $k\ll a\,H$, 
\begin{equation}
\mathcal{R}_k(N)\simeq \mathcal{R}_k(N_c)+\mathcal{R}_k'(N_c)\int_{N_c}^N\exp\bigg\{\int_{N_c}^{\hat N}\Big[2\eta(\tilde N)-3\Big]\diff \tilde{N}\bigg\}\diff \hat{N},
\label{eq:second_mode}
\end{equation}
where $N_c$ is some reference time which we may choose to be given by $k=a (N_c)H(N_c)$. In the SR limit, $\eta\simeq 0$ and the second term of Eq.\ (\ref{eq:second_mode}) decays exponentially, so $\mathcal{R}_k(N) \simeq \mathcal{R}_k(N_c)$. However, during the USR phase this solution grows, and the power spectrum is enhanced. During the subsequent CR phase with $\eta\lesssim 0$ the solution decays again, and the modes of $\mathcal{R}$ that contribute to the peak of the power spectrum freeze to their final values roughly at the end of the USR phase, see e.g.\ \cite{Ballesteros:2020sre}. From \eqref{eq:mukhanov-sasaki} (also from \eqref{eq:second_mode}), one can see that the monotony of $\mathcal{R}$ changes at $\eta = (3-\epsilon)/2\approx 3/2$. The value $\eta=3/2$ is achieved twice along the unperturbed trajectory: once in the SR-USR transition, and another time in the USR-CR transition. We denote as $N_*$ the time at which $\eta = 3/2$ along the unperturbed trajectory in the USR-CR transition.

\section{The $\delta N$ formalism}\label{s:classical}
\label{sec:deltan}

The $\delta N$ formalism \cite{Sugiyama:2012tj} allows to compute the curvature perturbation $\zeta$
non-linearly on large distances (compared to the Hubble scale) as a function of the flat-gauge, linear perturbations $\delta \phi$ and $\delta \pi$, using only the unperturbed equations of motion \eqref{eq:EoMphi}. This provides an explicit relation between the non-linear  $\zeta$ 
and the linear $\delta \phi$ and $\delta \pi$, which can be exploited to compute the probability distribution function (PDF) of the former provided one knows the PDF of the latter. We will now summarize the essence of the $\delta N$ formalism and we refer the reader to Appendix~\ref{ap:deltan} (and references therein) for technical details.

The $\delta N$ formalism crucially relies on the validity of a \emph{gradient expansion}, i.e. an expansion of the non-linear equations of motion in powers of gradients (in position space), or powers of $k/(aH)$ (in Fourier space). First, using such an expansion, one can write down a \emph{$\delta N$ formula} relating  $\zeta$
(on large scales) to the local expansion between two conveniently-chosen hypersurfaces (which are appropriately determined by the values of $\delta \phi$ and $\delta \pi$ on them). Second, from the gradient expansion follows the so-called \emph{separate universe approach} (see \cite{Wands:2000dp}), which states that non-linearly perturbed fields, coarse-grained on large scales, are described by differential equations that look like those governing the unperturbed (homogeneous) fields.
Invoking the separate universe approach, one can then relate the aforementioned local expansion to the number of $e$-folds computed using the unperturbed equations of motion.

In practice, the prescription is the following. We specify an initial hypersurface, determined by an $e$-fold time $N_i$ and unperturbed field and momentum values $\bar{\phi}(N_i)$, $\bar{\pi}(N_i)$.\footnote{We recall that $\bar{\phi}$ and $\bar{\pi}$ are solutions to \eqref{eq:EoMphi} with initial conditions fixed sufficiently before the USR phase, such that they have converged to the SR attractor before entering the USR phase of their dynamics (see comment below \eqref{eq:sr-pars}).} 
Ideally, in a calculation without approximations, the specific choice of $N_i$ would be immaterial. However, in a practical application of the $\delta N$ formalism such as the one we implement,  $N_i$ should be
chosen late enough such that the linear Fourier modes of 
$\zeta$ (or equivalently $\mathcal{R}$)
have frozen and gradients become negligible~\cite{Lyth:2004gb}.
At the same time, $\delta \phi$ behaves linearly only at early times when the kinetic terms dominate the dynamics,
so choosing $N_i$ too late would neglect non-linearities and therefore using the linear approximation to compute the variance of $\zeta$ would become less accurate.
As a consequence of the impossibility of optimizing both aspects
simultaneously, a
spurious dependence on $N_i$ remains. As we shall discuss in this section, the details of the phase space of the inflationary model determine how strongly the PDF of $\delta N$ actually ends up depending on the choice of $N_i$. In Sec.~\ref{sec:stochastic}, we will discuss the relation between the choice of $N_i$ and that of the coarse-graining parameter $\sigma$ in stochastic inflation. For a strongly peaked power spectra like the ones arising in the presence of an USR phase, the time of freezeout of the Fourier modes corresponding to the peak of the power spectrum provides a choice for $N_i$ yielding consistent results with the stochastic $\delta N$ formalism. 
Under certain circumstances,  
the dependence on $N_i$ of the PDF of  $\zeta$
may vanish completely, as we will discuss later. 

Second, we perturb $\bar{\phi}(N_i)$, $\bar{\pi}(N_i)$ by $\delta\phi(N_i)$ and $\delta\pi(N_i)$ (henceforth $\delta\phi$ and $\delta\pi$). Third, we compute the difference in the number of $e$-folds $\delta N$ needed to reach some later stage of inflation given by a fixed field value $\bar{\phi}(N_f)$ (e.g. the end of inflation) with respect to the unperturbed case by evolving the \emph{perturbed initial conditions} $\bar{\phi}(N_i)+\delta\phi$, $\bar{\pi}(N_i)+\delta\pi$ with the \emph{unperturbed equations of motion} \eqref{eq:EoMphi}. By virtue of the $\delta N$ formula, this difference $\delta N$ coincides with the (frozen) non-linear perturbation  $\zeta$.  
We thus reach a non-linear relation between $\delta N =\zeta$
and $\delta\phi$, $\delta\pi$. Formally: 
\begin{align} \nonumber 
\zeta(N_f)= \delta N(N_f) = &\,\, N\left[\{\bar{\phi}(N_i)+\delta\phi,\bar{\pi}(N_i)+\delta\pi\}\to \bar{\phi}(N_f) \right] \\ & -N\left[\{\bar{\phi}(N_i),\bar{\pi}(N_i)\}\to \bar{\phi}(N_f)  \right]\,, \label{ec:formula_deltaN} 
\end{align} 
where $N[\{\phi_1\,,\pi_1\}\to \phi_2 ]$ denotes the number of $e$-folds that it takes to reach the field value $\phi_2$ evolving the unperturbed equations of motion \eqref{eq:EoMphi} with the initial conditions $\{\phi_1\,,\pi_1\}$. The dependence on $N_f$ in \eqref{ec:formula_deltaN} vanishes because of the conservation of  $\zeta$
for $N>N_i$ \cite{Lyth:2004gb}. The residual dependence on $N_i$ will be discussed in the rest of this section.

As explained above, the $\delta N$ formalism provides a non-linear relation between $\delta N$ and the field perturbations $\delta \phi$, $\delta\pi$ at some $e$-fold time $N_i$, which can then be used to obtain the PDF of the non-linear perturbation $\zeta=\delta N$. 
Let us emphasize that the dynamics or statistics of $\delta\phi$ and $\delta\pi$ are not determined in this formalism through any action or equation of motion; in this context, $\delta\phi$ and $\delta\pi$ are just a set of variables whose functional relation to $\delta N$ we need to know in order to infer the statistical properties of the latter by sampling them. In general, $\delta N$ will depend non-linearly on both $\delta\phi$ and $\delta\pi$. Let us now consider for simplicity that $\delta\phi$ and $\delta\pi$ are {\it not} independent of each other (an assumption whose validity we will discuss later on). In this case, and if we know the PDF of $\delta\phi$, we can obtain the PDF of $\zeta$ via: 
\begin{equation}
P (\zeta) 
= P (\delta \phi ) 
\bigg|\frac{\diff \delta \phi}{\diff \delta N}\bigg|\,.
\label{changeprob}
\end{equation}
It is however important to stress that the $\delta N$ formalism does not provide any information a priori about the statistical properties of $\delta\phi$, which is an input that needs to be introduced by hand. If, as it is often done in the literature \cite{Atal:2019erb,
Biagetti:2021eep,Pi:2022ysn,Tomberg:2023kli,Hooshangi:2023kss}, the probability density function (PDF) of $\delta\phi$ is assumed to be Gaussian
\begin{equation}\label{ec:pdfdeltaphi}
P ({\delta\phi}) = \frac{1}{\sigma_{\delta\phi}\sqrt{2\pi}}
\exp\bigg(-\frac{\delta\phi^2}{2\sigma_{\delta\phi}^2}\bigg)\,,
\end{equation}
with variance $\sigma_{\delta \phi}^2$, the non-linear relation between $\delta N$ and $\delta\phi$ given by the $\delta N$ formalism results in a non-Gaussian PDF for $\zeta$
through Eq.\ \eq{changeprob}. We will assume Gaussianity of $\delta\phi$ in this work for the sake of simplicity, but we emphasize that obtaining the correct PDF of $\zeta$
requires, in principle, full knowledge of the statistics of inflation fluctuations.  Determining if $\delta\phi$ is Gaussian away from the maximum of the PDF is, in general, not a simple task.\footnote{For instance, the bispectrum in a model similar to our {\it Model 1} was discussed in \cite{Taoso:2021uvl} using the in-in formalism.
For smooth transitions, the final bispectrum of $\mathcal{R}$ in that case is of the local form 
and its magnitude is controlled by the value of $\eta$ in the CR phase. A consistent result can be obtained by expanding the $\delta N$ result to second order in perturbations, see \cite{Pi:2022ysn}.}
The use of perturbation theory to compute correlators around the maximum of the PDF does not seem a practical strategy to find the statistical properties of large fluctuations
because an extrapolation is in general not well justified. For this reason, some non-perturbative approaches have recently started to be considered, see in particular \cite{Celoria:2021vjw}.

\subsection{Approximate solutions in classical $\delta N$}\label{ss:UAT}

In this section, we
discuss two different (but related) approximate implementations of the $\delta N$ formalism to compute $\delta N(\delta\phi, \delta\pi)$. Both approaches relate $\delta\phi$ and $\delta N$ according to
\begin{equation}
\delta\phi = \bar\phi(N_i-\delta N)-\bar\phi(N_i)\,,
\label{eq:delta_n_one}
\end{equation}
where $\bar\phi$ evolves according to \eqref{eq:EoMphi} and the initial time $N_i$ is chosen after the {\it linear} $\zeta$, or equivalently $\mathcal{R}$,
has become constant, as mentioned earlier. The importance of the non-linear relation between $\delta\phi$ and $\zeta$
is manifest in this expression. As we shall see in Eq.\ \eqref{eq:deltaphiCR}, solving Eq.\ \eqref{eq:delta_n_one} for constant $\eta$ and Taylor-expanding the result around $\delta N=0$, we obtain
\begin{equation}\label{ec:leading_order}
\delta\phi=-\pi\,\delta N+\mathcal{O}(\delta N^2)\,,
\end{equation}
where the equality has to be evaluated at a time $N_i$ after (the linear) $\mathcal{R}$ has become constant. The leading order in \eqref{ec:leading_order} is nothing but the standard linear relation between $\zeta$
and $\delta\phi$. The non-linear terms $\mathcal{O}(\delta N^2)$ are responsible for the non-Gaussian behavior in $\zeta$
for large values of the fluctuations, even if $\delta\phi$ is assumed to be fully Gaussian, as discussed above.

Notice that the formula \eq{eq:delta_n_one} implicitly fixes $\delta\pi$ as a function of $\delta\phi$ in such a way that $\bar{\pi}(N_i)+\delta\pi$ lies on the unperturbed trajectory of the inflaton in phase space. There is, in principle, no reason why this (implicit) choice of $\delta\pi$ should be the right one. The prescription has nevertheless been used in the literature, see e.g.\ \cite{Garriga:2015tea,Hooshangi:2023kss}, to obtain analytical results. 

\subsubsection{Eternal CR approximation}

Let us assume that the field $\phi$ moves along its unperturbed trajectory in phase space, according to Eq.\ \eqref{eq:delta_n_one}. In addition, we {\it assume} that this trajectory
describes a CR phase. We will call this the {\it eternal CR approximation}. As we will see, this approximation always works well if very small fluctuations are considered. We will also see (later in Section \ref{ss:full}) that if this approximation is used for larger fluctuations it can work well (or not) depending on the shape of the phase space. Physically, this corresponds to an inflaton released close to the top of a quadratic potential in the distant past.

Along a CR trajectory,  $\bar\phi(N) \sim e^{-\eta N}$ (where $\eta$ is constant) and the homogeneous perturbations $\delta\phi$ (along the unperturbed trajectory) obey~\cite{Tomberg:2023kli}
\begin{equation}
\delta \phi
=
\left(e^{\eta\,\delta N} - 1\right){\bar\phi(N_i)}\,.
\label{eq:deltaphiCR}
\end{equation}
Denoting the {\it linear} power spectrum of $\zeta$ as 
\begin{equation}
\mathcal{P}_\mathcal{\zeta}(k) \equiv \frac{k^3}{2\pi^2}|\mathcal{\zeta}_k|^2\,,
\end{equation}
evaluated once the modes are super-Hubble and have become constant, and using that\footnote{\label{ft:clarification_zeta} We emphasize that $\mathcal{P}_\mathcal{\zeta}$ is the power spectrum of $\zeta$ computed \emph{in linear perturbation theory}, which coincides with the linear $\mathcal{P}_\mathcal{R}$, and therefore can be computed by solving \eqref{eq:mukhanov-sasaki}. Correspondingly, $\sigma^2_\zeta$ is the variance of the linear (Gaussian) $\zeta$, which does not coincide (in general) with the variance of the non-linear $\zeta$ (whose full PDF we intend to compute).}
\begin{equation}
\sigma_{\delta\phi}^2 = \int \mathcal{P}_{\delta\phi} (k) \, \diff \log k
= \bar\pi(N_i)^2 \int  \mathcal{P}_{\zeta} (k) \diff \log k \equiv \bar\pi(N_i)^2  \sigma_{\zeta}^2\,,
\label{eq:varphi}
\end{equation}
the PDF of $\zeta=\delta N$
turns out to be \cite{Tomberg:2023kli, Pi:2022ysn} 
\begin{equation}\label{ec:pdeltaNfull}
P(\zeta)=\frac{1}{\sqrt{2\pi\, \sigma^2_{\zeta}}}\exp\left[-\frac{1}{2\sigma^2_{\zeta}\,\eta^2}\left(1-e^{\eta\,\delta N}\right)^2+\eta\,\delta N\right]\,.
\end{equation}
In the limit of large $\zeta$,
its PDF scales as 
\begin{align} \label{eexp}
P(\zeta) \sim e^{\eta\zeta}.
\end{align} 
We point out that the dependence on $N_i$ of the PDF of $\delta N$ that might have come from $\bar\pi(N_i)^2$ in \eq{eq:varphi} actually vanishes identically. However, there is in principle a residual dependence on $N_i$ that is not explicit in the above expressions. It comes from the definition of the linear $\sigma^2_{\mathcal{\zeta}}$; in particular from the upper limit of integration. 

\subsubsection{Unperturbed-trajectory approximation}

\begin{figure}
\centering
\includegraphics[width=0.49 \textwidth]{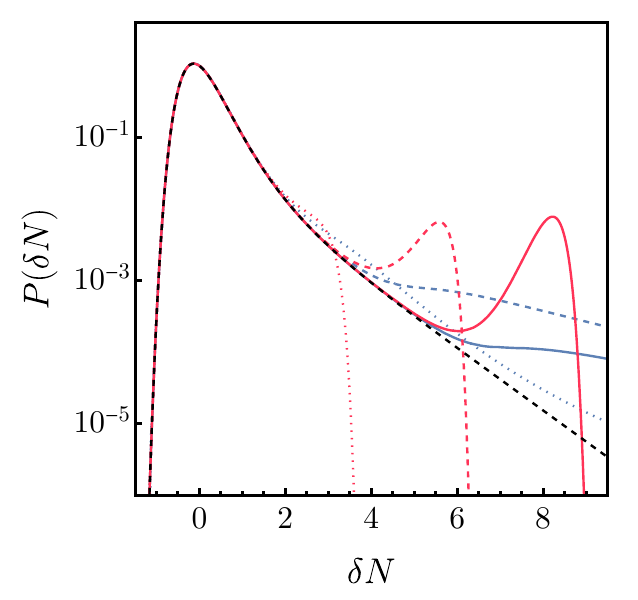} 
\includegraphics[width=0.49 \textwidth]{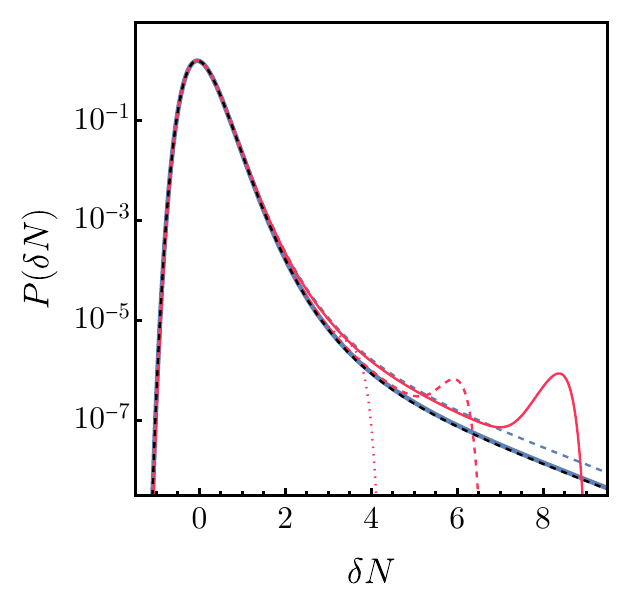} 
\caption{
\it Left panel: Probability density for $\delta N$ for {\it Model 1}. The black dashed-line corresponds to the eternal CR approximation in Eq.\,(\ref{eq:deltaphiCR}), with $\eta$ in the CR phase following USR. The red lines correspond to a numerical evaluation of Eq.\,(\ref{eq:delta_n_one}) for $N_i-N_{*}=2$ (dotted), $4$ (dashed), and $6$ (solid). The blue lines correspond to the solutions obtained by including the momentum perturbations according to Eq.\,(\ref{eq:2shift}) for the same values of $N_i-N_{*}$. The dependence of the PDF on the choice of $N_i$ is evident. Right panel: Same as left panel but for {\it Model 2}.
The red lines correspond to $N_i-N_{*}=3$ (dotted), $5$ (dashed), and $7$ (solid). The blue lines correspond to $N_i-N_{*}=3$ (solid) and $7$ (dashed).
}
\label{fig:pdf}
\end{figure}

Let us now consider a model of USR inflation where the subsequent CR phase has a finite duration in the past. Thus, a sufficiently large $\delta\phi$ perturbation makes $\bar{\phi}+\delta\phi$ fall outside of that CR phase and into the USR part of the unperturbed trajectory (with $\eta$ changing from negative to positive).\footnote{This approximation and the eternal CR one discussed above coincide form sufficiently small fluctuations, since we are considering models in which USR is followed by a constant CR phase of finite duration.} Again, the solution for $\delta N(\delta\phi)$ can be obtained by numerically solving \eqref{eq:delta_n_one}. This \emph{unperturbed-trajectory approximation} then leads to a bump in the PDF of $\zeta=\delta N$ located at a large enough value of $\delta N$ and superimposed on the behaviour described by Eq.\ \eqref{eexp}. The emergence of such a (spurious) bump can be understood from \eqref{changeprob}. As seen there, the PDF of $\delta N$, $P(\delta N)$ is the product of $P(\delta\phi)$ and the absolute value of the Jacobian of the change of variables. While $\bar{\phi}+\delta\phi$ falls inside the CR region (where $\eta<0$), the latter provides the dominant contribution to $P(\delta N)$ for large $\delta N$ (cf.\ Eq.\ \eqref{eexp}), which corresponds to an exponential decay. Now, if $\delta\phi$ is such that $\bar{\phi}+\delta\phi$ falls into the USR part of the unperturbed trajectory (as it will eventually happen in the unperturbed-trajectory approximation), the same functional form for $P(\delta\phi)$ and the Jacobian still holds, with the sole addition of some constant factors that ensure the continuity of $\delta N$ across the CR-USR transition (that in any case do not affect the behaviour of $P(\delta\phi)$ or the Jacobian).
In the USR phase, $\eta>0$, and therefore the Jacobian term grows as $\exp(\eta\,\delta N)$, while $P(\delta\phi)$ provides a superexponential suppression for large $\delta N$ (via the term $\exp\{-[1-\exp(\eta\,\delta N)]^2\}$ in \eqref{ec:pdeltaNfull}). The competition between these two terms (dominated by the former initially, and by the latter for large $\delta N$) yields a bump followed by a violent (superexponential) cutoff. {Moreover, this ultimately induces a dependence on $N_i$ in the PDF of $\delta N$, due to the larger (or smaller) values of $\delta\phi$ needed for $\bar{\phi}+\delta\phi$ to probe the USR phase of the unperturbed trajectory if $N_i$ is chosen later (or earlier) on in the CR phase}. This {behaviour} is illustrated in the red curves of Fig.\,\ref{fig:pdf} for the two models under consideration, together with the PDF resulting from the eternal CR approximation in Eq.\,(\ref{eq:deltaphiCR}).
As we will show shortly, the spurious bump in the PDF disappears when momentum perturbations are appropriately accounted for. Interestingly, once these are included, the result for the PDF of $\delta N$ obtained is actually closer to the one obtained by \emph{unrealistically} assuming the CR phase of the unperturbed trajectory to extend indefinitely to the past of $N_i$, although significant deviations from it can arise depending on the specific shape of the phase space attractor close to the local maximum of the potential, as we shall discuss.

\subsection{Full calculation}\label{ss:full}

Let us now consider the full calculation of $\delta N$,	 consistently accounting for both field and momentum perturbations without making any approximation. As we mentioned earlier, $N_i$ must be chosen as a time posterior to the time at which the linear $\mathcal{R}=-\delta\phi/\bar{\pi}$ (or linear $\zeta$) becomes constant. Differentiating this last expression, we obtain (see also \cite{Tomberg:2022mkt}) the following relation between the fluctuations $\delta\pi$ and $\delta\phi$:
\begin{equation}
\label{eq:2shift}
\delta \pi  = \frac{\bar{\pi}^\prime(N_i)}{\bar{\phi}^\prime(N_i)} \delta \phi\,.
\end{equation}
In other words, the freezeout condition for linear perturbations (which is necessary to apply consistently the $\delta N$ formalism) implies a relation between $\delta\phi$ and $\delta\pi$, which allows to still use \eqref{changeprob} to compute the PDF of (the non-linear) $\zeta=\delta N$ using the PDF of $\delta\phi$ as sole input. Yet, this relation is not, in general, the one implicitly imposed assuming \eqref{eq:delta_n_one}. To compute $\delta N$, one simply needs to solve Eq.\ \eq{eq:EoMphi} scanning through the initial conditions consistently with \eqref{eq:2shift}, in order to sample the PDF. Specifically, once we have determined $N_i$ via the peak freezing condition --see the discussion above Eq.\ \eq{ec:formula_deltaN}--, we perturb $\bar{\phi}(N_i)$ by some value $\delta\phi$, and $\bar{\pi}(N_i)$ by the $\delta\pi$ given by \eqref{eq:2shift}. The corresponding $\delta N$ is the difference in the number of realized $e$-folds of inflation between the shifted trajectory and the unperturbed one. In this way, one obtains a relation $\delta N[\delta\phi,\delta\pi(\delta\phi)]=\mathcal{\zeta}(\delta \phi)$,
and assuming $\delta\phi$ to be Gaussian, one can again use \eqref{changeprob} to obtain the PDF of $\delta N=\zeta$.
The PDF resulting from applying this prescription to the models under consideration is depicted in the blue curves of Fig.\,\ref{fig:pdf}.

\begin{figure}
\centering
\includegraphics[width=0.49 \textwidth]{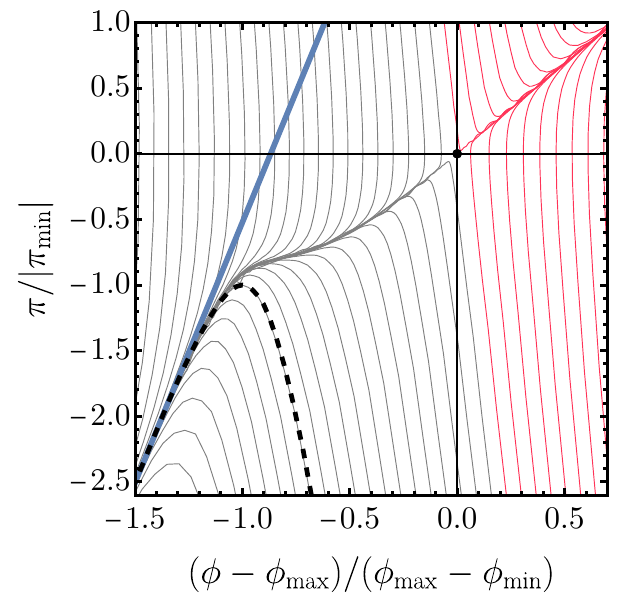} 
\includegraphics[width=0.49 \textwidth]{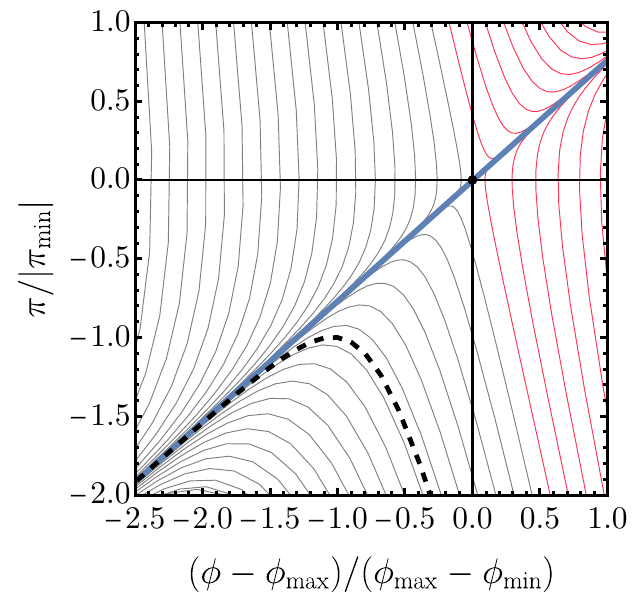} 
\caption{
\it Left panel: Phase space for {\it Model 1}. For red trajectories, the field turns back before reaching the local maximum and gets stuck in the local minimum.  The black dot represents the unstable equilibrium point $\phi_{\rm max}$ at which the field sits in the local maximum with zero velocity. The black, dashed line represents the unperturbed trajectory, and the blue line denotes the CR line. We use $\pi_{\rm min}$ to denote the minimum velocity reached in the unperturbed trajectory, and $\phi_{\rm min}$ denotes the corresponding field value. Right panel: Same as left panel for {\it Model 2}.}
\label{fig:phase}
\end{figure}

It is instructive to analyse the structure of the phase space for both of our models, depicted in Fig.\,\ref{fig:phase}.\footnote{For the model in Eq.\,(\ref{eq:step_model}) we reconstruct the potential as a function of the field using Eq.\,(\ref{eq:pot_slowroll}), and then solve the Klein-Gordon equation with different initial conditions using the resulting $V[N(\phi)]$.} The tangent to any trajectory in phase space (in particular, the unperturbed trajectory) is given by
\begin{equation} \label{att}
\frac{\diff \pi}{\diff \phi} =\epsilon-\eta \simeq -\eta\,.
\end{equation}
For field values smaller than the local maximum of the potential,\footnote{We recall that we assume (without loss of generality) that the field decreases as inflation proceeds.} the trajectories of the system are driven to an attractor, see Fig.\ \ref{fig:phase}. In the thin grey trajectories, the field gets past the local maximum of the potential and reaches its absolute minimum, whereas for the thin red trajectories it gets stuck in the local minimum, leading to eternal inflation. Notice how the grey trajectories converge to an attractor for $\phi<\phi_{\max}$ (the local maximum of the potential), which, as it is to be expected, at some point overlaps with the unperturbed trajectory (black dashed). The tangent to this attractor, given by \eqref{att}, reaches a phase of constant slope (i.e.\ the attractor is a CR trajectory). This occurs almost immediately to the left of $\phi_{\max}$ for \textit{Model 2} (right panel), but not for \textit{Model 1} (left panel), where the $\eta$ of the attractor varies rapidly before stabilizing. The existence of a region of non-constant slope of the attractor in {\it Model 1} close to the local maximum of the potential can be traced back to $\eta_V(\phi)$ varying rapidly near that point.\footnote{In the right panel, the opposite occurs: $\eta_V(\phi)$ is close-to-constant (i.e.\ the potential is well described by an inverted parabola) around the local maximum, yielding a close-to-straight (i.e.\ CR) attractor right after the local maximum.} The blue line represents the \emph{CR line}, whose tangent vector is given by (\ref{att}) evaluated at $N_i$ (or equivalently $N>N_i$, since $\eta$ is already constant along the unperturbed trajectory at that point). The points along this line indicate the initial conditions of a perturbed trajectory $\{\bar{\phi}(N_i)+\delta\phi, \bar{\pi}(N_i)+\delta\pi\}$ consistent with \eqref{eq:2shift}. Choosing $N_i$ when the modes around the peak of the power spectrum freeze-- toward the beginning of the CR phase-- corresponds to fixing $\{\bar{\phi}(N_i), \bar{\pi}(N_i)\}$  in the bottom left corner of both panels of Fig.~\ref{fig:phase}.

\subsection{Comparison between the full and the approximate implementations of $\delta N$}

Fig.\ \ref{fig:phase} illustrates the difference between the unperturbed-trajectory approximation and the full calculation. If momentum perturbations $\delta\pi$ were chosen as in the unperturbed-trajectory approximation (Sec.\ \ref{ss:UAT}), the field would be shifted at $N_i$ to some point in the black dashed line, and evolve hereon along said line. Instead, accounting for the momentum perturbations as in \eqref{eq:2shift}, the field is shifted at $N_i$ to some point on the blue line, and evolves hereon along the corresponding grey trajectory. For small enough perturbations (that is, for sufficiently small $\delta\phi$, or $\delta N$), the field shifted at $N_i$ remains on the overlapping locus of the CR attractor and the unperturbed trajectory, and both approaches are essentially equivalent. This is consistent with Fig.\,\ref{fig:pdf}, as the PDF for $\delta N$ obtained through all the different approaches agree sufficiently close to the peak of the distribution. Deviations from the unperturbed-trajectory $\delta N$ formalism due to the inclusion of $\delta\pi$ satisfying \eqref{eq:2shift} only appear for fluctuations that are large enough for the trajectory of the perturbed field to become sensitive to the region of phase space beyond the turnaround of the unperturbed trajectory, where the CR line (solid blue) and the unperturbed trajectory (black dashed) coincide no more. Finally, notice that when momentum perturbations are chosen consistently with \eqref{eq:2shift}, values of $\delta N\to\infty$ are sampled for $\delta\phi\to \bar{\phi}(N_i)-\phi_{\max}$ (more specifically, for the $(\delta\phi,\delta\pi)$ at which the blue line crosses from the grey region to the red region in Fig.\ \ref{fig:phase}). Conversely, in the unperturbed-trajectory approximation, $\delta N\to\infty$ is only achieved as $\delta\phi\to \infty$ along the unperturbed trajectory of the field. 

One can notice in Fig.\,\ref{fig:pdf} that the PDF for {\it  Model 2}  obtained by sampling $\delta\phi$ and $\delta\pi$ consistently with \eqref{eq:2shift} is remarkably close to the one obtained using the eternal CR approximation in Eq.\,(\ref{eq:deltaphiCR}) for a wide range of $N_i$ values, and only begins to deviate for very late (and therefore inadequately chosen) values of $N_i$, long after the modes corresponding to the peak of the power spectrum have frozen. The reason for this behaviour is that the eternal CR approximation effectively corresponds to shifting the field at $N_i$ along the CR line (with tangent vector given by \eqref{att} evaluated at $N_i$) and assuming it evolves hereon along the same CR line. This approximation is indeed an accurate description of the actual dynamics of \textit{Model 2}, shown in the right panel of Fig.\,\ref{fig:phase}, where the CR line (solid blue) and the phase space attractor (convergence of grey lines) lie on top of each other. However, this is not necessarily the case for other implementations of USR, and in fact it does not occur for {\it Model 1}, as shown in the left panel of the same figure. In that case, the field originally shifted along the CR line does not evolve afterwards along it, but along the grey curves that intersect it. This translates into notable differences in the deep-tail regime when comparing the PDF obtained with the eternal CR approximation and the one obtained by including the momentum perturbations as per \eqref{eq:2shift}, as seen in Fig.\ \ref{fig:pdf}. 

\section{Stochastic inflation}\label{s:stoch}
\label{sec:stochastic}

The framework of stochastic inflation~\cite{1986LNP...246..107S} splits the inflaton into short and long-wavelength Fourier modes in order to study the backreaction of the former on the latter.
The modes with small wavelength
evolve as free fields, relatively insensitive to the non-linearities induced by the interaction terms in the scalar potential. Hence, it is assumed that the evolution of these short modes is well captured by linear, deterministic, equations of motion. Their effect on the long modes is then accounted for by stochastic source terms which are assumed to behave as white noise, due to the underlying Gaussian statistics assumed for the short modes and the choice of the coarse-graining window that separates short and long modes (see \cite{Vennin:2020kng} for a review). This separation is done introducing a (small) coarse-graining parameter $\sigma$
such that a mode of wavelength $k$ 
is considered a long one at any given time provided its wavenumber is smaller than a quantity 
\begin{align}
k_{\sigma}\equiv\sigma aH\,.
\end{align}
In the presence of an USR phase, $\sigma$ must be chosen small enough in such a way that the modes around the peak of the power spectrum\footnote{These modes provide the largest kicks, as it will become clear in \eqref{eq:stoch_force1}.} have had enough time to freeze out and classicalize when they become part of the coarse-grained field\footnote{The modes around the peak (which undergo the most superhorizon enhancement due to USR) must not become part of the coarse-grained field until they freeze, which occurs slightly after the end of USR. Thus, if such a mode takes $\Delta N_{\rm freeze}$ $e$-folds to freeze, $\sigma$ should be, at least, $e^{-\Delta N_{\rm freeze}}$. Later on we will illustrate this in an example, and explain how it relates to the choice of $N_i$ in classical $\delta N$.} \cite{Ballesteros:2020sre} (see \cite{Jackson:2023obv} for a recent discussion and \cite{Launay:2024qsm} for a formulation of stochastic inflation beyond the separate universe approach). However, it has been argued that $\sigma$ cannot be chosen arbitrarily small, as it would result in the loss of information about the non-linear evolution of the curvature perturbation \cite{Figueroa:2020jkf}. There is therefore a tradeoff between choosing $\sigma$ too small and too large. In the following, we use $\sigma=0.1$ and $\sigma=0.01$ as two representative values for the USR phases of Models 1 and 2 with duration $\Delta N \approx 2.5\approx -\log(0.08)$.

Once the quantum fluctuations of the field freeze out they can be treated as classical random variables \cite{Ballesteros:2020sre}. These fluctuations backreact on the background trajectory, leading to a system of stochastic differential equations which reads
\begin{align}
\label{eq:stochasticinflation_first}
\diff \phi &= \pi \diff N + \diff W_\phi\,, \\
\diff \pi &= -(3-\epsilon) (  \pi + \partial_\phi\log V) \diff N + \diff W_\pi\,,
\label{eq:stochasticinflation}
\end{align}
where $\phi$ and $\pi$ are now stochastic variables and $\diff W_{\phi,\pi}$ denote Wiener increments, often expressed in terms of stochastic ``kicks'' $ \xi_{\phi,\pi} =\diff W_{\phi,\pi} / \diff N$ satisfying
\begin{align}
\langle\xi_{\phi}(N)\xi_{\phi}(N')\rangle = \mathcal{P}_{\delta\phi}(N,\kc)\delta(N-N')\,,\\
\langle\xi_{\pi}(N)\xi_{\pi}(N')\rangle = \mathcal{P}_{\delta\pi}(N,\kc)\delta(N-N')\,.
\label{eq:varxi}
\end{align}
In what follows we assume that the power spectra appearing in the variance of these noises is computed solving (\ref{eq:mukhanov-sasaki}) with the coefficients evaluated in the background trajectory. In principle, a full treatment of stochastic inflation would involve using the stochastic variables in the coefficients and recalculating the power spectrum at every time step, as in \cite{Figueroa:2021zah}. As the results of \cite{Figueroa:2021zah} imply, however, not only does this require plenty of computational power, but, more importantly, does not seem to have any effect on the final result for the probability distribution. Using the background quantities for the coefficients it is therefore a practical working approximation which should provide excellent accuracy. We will comment on the reasons for this behaviour at the end of this section.

Despite having two seemingly independent Wiener increments in the equations of motion, there is actually only one independent kick at each time step. This is due to the fact that the quantum state of the short-wavelength perturbations is highly squeezed~\cite{Grain:2019vnq,Figueroa:2021zah,Tomberg:2022mkt,
Tomberg:2023kli}, which forces the kicks to be aligned in a very specific way, with $\xi_\pi = (\delta\pi_k / \delta\phi_k )\xi_\phi$. Once the system enters the last CR phase, the (linear) curvature perturbation freezes to a time-independent value on super-horizon scales. By differentiating the linear expression {$\mathcal{R}=-\delta\phi/\bar\pi$}, we obtain the relation\footnote{Cf.\ \eqref{eq:2shift}, where an analogous relation for field/momentum perturbations holds.}
\begin{equation}
    \dfrac{\xi_\pi}{\xi_\phi} = \dfrac{\overline \pi'}{\overline \phi'}\,,
    \label{eq:noiserelation}
\end{equation}
where the quantities carrying a bar over are evaluated on the unperturbed trajectory, and are therefore deterministic. The variance of the Wiener increment at each time step can be expressed as (for the case of $\phi$ and analogously for $\pi$):
\begin{equation}
\label{eq:stoch_force1}
\langle \diff W_\phi^2 \rangle = \overline{\pi}(N)^2 \mathcal{P}_{\cal \zeta}(k_\sigma)\frac{\diff \log k_\sigma}{\diff N} \diff N\,,
\end{equation}
where $\mathcal{P}_{\zeta}$ represents the linear power spectrum of $\zeta=\mathcal{R}$ after the modes have frozen (recall Footnote~\ref{ft:clarification_zeta}).
Since the constant parameter $\sigma \ll 1$ is $k$-independent, the field receives a kick from a single scale at every time step.\footnote{See \cite{Jackson:2023obv} for a recent proposal on the topic.} By differentiating $k_\sigma$, we find $\diff \log k_\sigma /\diff N = 1-\epsilon \simeq 1$. We remark that the factor $\overline{\pi}(N)^2$ denotes the momentum of the unperturbed background field (i.e.\,the solution to eqs.\,(\ref{eq:stochasticinflation_first}, \ref{eq:stochasticinflation}) without the stochastic noise terms), and it takes into account the time dependence of the power spectrum $\mathcal{P}_{\delta\phi}(N,k) = \overline{\pi}(N)^2  \mathcal{P}_{\zeta}(k)$ where, if $\sigma$ has been chosen adequately, we can assume that $\mathcal{P}_{\cal \zeta}(k)$ has settled to its frozen value.

The resulting shift in the duration of inflation $\delta N$ induced by the stochastic kicks can be extracted at late times, once the peak modes have frozen out (and hence have become coarse-grained), and therefore subsequent kicks become inefficient. This can be accomplished in practice by turning off the noise terms in the stochastic equations once they become irrelevant in comparison to the drift term, and integrating the hereafter unperturbed (i.e.\ noiseless) system until inflation ends (that is, until $\epsilon=1$). The value of $\delta N$ can then be extracted by comparing the duration of inflation in this realization to that of the unperturbed trajectory. We remark that it is important to integrate the system until the end of inflation since, even if the noise is turned off, the field may not be in the unperturbed trajectory (or even in the phase-space attractor) if integration is cut short. In the limit in which the linear power spectrum of $\zeta$ is strongly peaked at some particular scale, the field receives the most significant stochastic kicks over a reduced amount of time. This corresponds to a single (possibly non-Gaussian) kick in the extreme case of a Dirac delta spectrum. The latter case is exactly equivalent to the standard $\delta N$ calculation of the previous section. This can be checked numerically by manually adjusting the width of the spectrum. It is therefore reasonable to expect both formalisms to yield very similar results for inflection-point models leading to highly peaked spectra. 

This discussion also makes clear that there is an approximate one-to-one correspondence between the parameter $\sigma$ in stochastic inflation and the parameter $N_i$ that denotes the time at which the field is perturbed in the classical $\delta N$ formalism. Indeed, for both results to match, $N_i$ must be chosen roughly as the time at which the peak in the power spectrum is coarse-grained, $k_{\rm peak}\simeq\sigma a(N_i)H(N_i)$. We reiterate, however, that this correspondence is only valid in the limit in which the power spectrum is highly peaked.

As in the classical $\delta N$ formalism, one can attempt to simplify the system of equations by assuming that the velocity of the scalar field adheres to the background trajectory at every time step, as suggested in \cite{Tomberg:2022mkt,Tomberg:2023kli}. This would indeed be the case if the stochastic kicks never took the field far from the phase space attractor. As we saw in the previous section for the case of the classical $\delta N$ formalism, this is a good approximation close to the peak of the PDF of $\delta N$. Whether it is a good approximation in the deep tail, however, strongly depends on the morphology of the phase space, and in particular, the variation of the slope of the attractor for $\phi<\phi_{\max}$. The corresponding equation implementing this approximation reads 
\begin{equation}
\label{eq:stochastic1}
\diff \phi = \overline{\pi}_c(\phi) \diff N + \diff W_\phi,
\end{equation}
where we have used $\overline{\pi}_c$ to denote the momentum of the unperturbed trajectory written as a function of the field value (and {\it not} as a function of $N$).\footnote{To be crystal clear, one should solve the background equation to determine $\overline{\pi}(N)$ and $\overline{\phi}(N)$, invert the latter to obtain $\overline{\pi}[N(\overline{\phi})]$, and then make the substitution $\overline{\phi}\to\phi$ in this deterministic equation to obtain $\overline{\pi}_c=\overline{\pi}[N(\phi)]$, which is now a stochastic variable.} This condition ensures that the scalar field stays on the unperturbed trajectory, denoted with a black dashed line in Fig.\,\ref{fig:phase}, at every time step. Let us emphasize that, in this scheme, the amplitude of the stochastic kicks in Eq.\,(\ref{eq:stoch_force1}) is still determined by $\overline{\pi}(N)$, and not $\overline{\pi}_c(\phi)$. The former is a deterministic quantity, whereas the latter is a stochastic variable due to its dependence on the stochastic variable $\phi$. Indeed, if the stochastic term were evaluated using $\overline{\pi}_c(\phi)$ instead of $\overline{\pi}(N)$, the induced stochastic time shift $\delta N$ would be necessarily Gaussian. This statement can be made clearer by transforming Eq.\,(\ref{eq:stochastic1}) into a stochastic equation for the time shift $\delta N$ \cite{Tomberg:2022mkt,
Tomberg:2023kli}. This equation can be expressed as
\begin{equation}
\label{eq:stochasticN}
\diff \delta N = -\frac{\overline{\pi}^\prime(N)}{2\overline{\pi}(N)} \mathcal{P}_{\zeta}(k_\sigma) (1-\epsilon)^2 \diff N +  \diff W_{\delta N},
\end{equation}
where the first term follows from Ito's lemma, and
\begin{equation}
\langle\diff  W_{\delta N}^2 \rangle = \frac{\langle \diff W_\phi^2 \rangle}{\overline{\pi}_c^2(\phi)}.
\end{equation}
It is clear from Eq.\,(\ref{eq:stoch_force1}) and Eq.\,(\ref{eq:stochasticN}) that replacing the momentum $\overline{\pi}(N)$ of Eq.\,(\ref{eq:stoch_force1}) by $\overline{\pi}_c(\phi)$ would entirely remove the path-dependence from the stochastic process and lead to a Gaussian distribution for $\delta N$. 

\begin{figure}
\centering
\includegraphics[width=0.49 \textwidth]{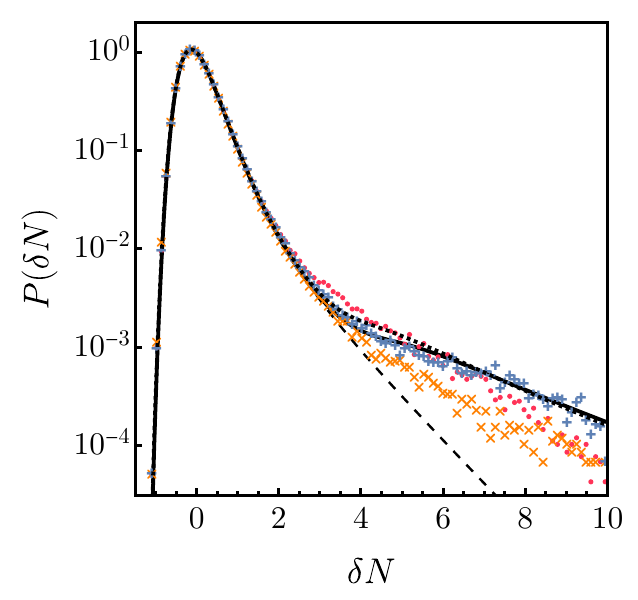} 
\includegraphics[width=0.49 \textwidth]{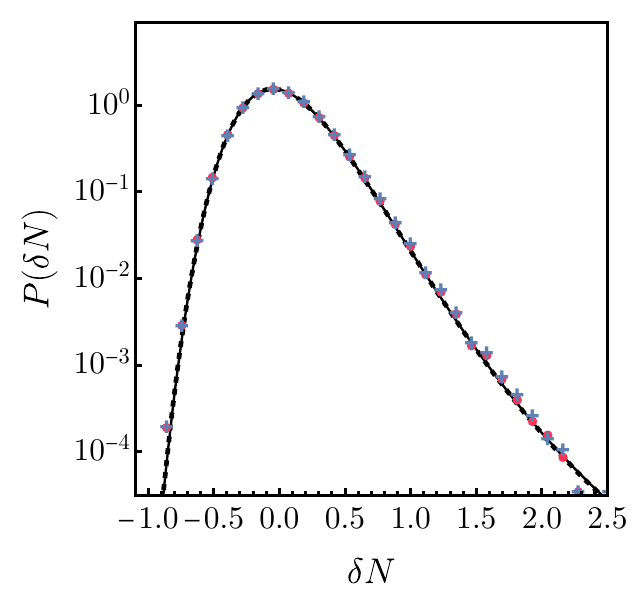} 
\caption{
\it Left panel: Probability density function for the model of Eq.\,(\ref{eq:step_model}). The black dashed-line denotes the eternal CR approximation. The black solid line is obtained by using the $\delta N$ formalism including the momentum perturbations, with $N_i-N_{*}=4.2$. The blue and orange crosses represent $10^6$ stochastic realizations of the system of two Langevin equations with $\sigma=10^{-2}$ and $10^{-1}$, respectively, and the red dots correspond to realizations of the simplified Langevin equation with $\sigma=10^{-2}$.{The black dotted line shows the PDF obtained by using the alternative procedure described around Eq.~\eqref{eq:noisevector2} to determine the direction of the kicks.} Right panel: Same as left panel for the model in Eq.\,(\ref{eq:poly_model}) with $\sigma=10^{-2}$.}
\label{fig:stoch}
\end{figure}

Let us now return to the case of interest for us, which is the solution to the full system of Eqs.\,(\ref{eq:stochasticinflation_first}, \ref{eq:stochasticinflation}). In the stochastic approach, the trajectory in phase space follows a random walk dictated by these equations, which can be recast as
\begin{align}
\begin{pmatrix}
\phi'\\\pi'
\end{pmatrix}
=\mathbfcal{D}+\begin{pmatrix}
\xi_\phi\\\xi_\pi
\end{pmatrix}\,,
\end{align}
where $\mathbfcal{D}$ is called the drift vector and (see Eqs.\ \eqref{eq:noiserelation} and \eq{att})
\begin{equation}	
\begin{pmatrix}\xi_\phi\\\xi_\pi\end{pmatrix} \propto \hat \xi \begin{pmatrix}1\\-\eta \end{pmatrix} \,,
\label{eq:noisevector}
\end{equation}
where $\hat \xi$ is a Gaussian-distributed noise and $\eta(N)$ is evaluated on the background trajectory at a given time $N$. Eq.\ \eqref{eq:noisevector} makes manifest that the CR line (blue line in Fig.\ \ref{fig:phase}) gives the direction of the stochastic kicks corresponding to the modes that are coarse-grained at $N_i$, see Sec.\ \ref{ss:full}.\footnote{These modes are the ones providing the \emph{larger} kicks, as $N_i$ corresponds to the freezing-out of the modes around the peak of the power spectrum. } Also, the drift vector $\mathbfcal{D}$ is tangent to the trajectories in phase space (grey lines in Fig.\ \ref{fig:phase}). If the attractor trajectory coincides with the CR line (as it happens for {\it Model~2}), the field remains on this line, kicked back and forth by the noise and driven along it by the drift. The system is therefore effectively described by a single degree of freedom. In this scenario, the dependence on $\sigma$ is lost; in other words, it does not matter when specifically the kicks are applied, but only their accumulated effect, as already noticed in \cite{Tomberg:2023kli}. This can be understood by analytically solving the single Langevin equation that governs the inflaton field constrained to drift and diffuse along the superposition of the CR line and CR attractor with constant $\eta$:
\begin{equation}\label{ec:ad0}
\phi' = -\eta\,\phi + \sqrt{2\epsilon(N)\mathcal{P}_\mathcal{\zeta}(k_\sigma)}\xi_\phi\,, 
\end{equation}
with $\epsilon(N)$ evaluated on the background trajectory at a time $N$. %
The solution is a Gaussian $\phi(N)=\bar{\phi}(N)+\delta\phi(N)$ with
\begin{equation}\label{ec:ad1}
\langle\phi(N)\rangle=\bar{\phi}(N) \, ,
\end{equation}  
and
\begin{equation}\label{ec:ad2}
\sigma^2_{\delta\phi}=2\epsilon(N)\int_{N_{\rm IR}}^N \diff N'\,\mathcal{P}_\zeta(k_\sigma(N')) = 2\epsilon(N)\int_{\log k_{\rm IR}}^{\log k_{\sigma}(N)}\diff \log k\,\mathcal{P}_\zeta(k)\,.
\end{equation}
If evaluated at $N_i$, the Gaussian $\phi(N_i)$ described by \eqref{ec:ad1}, \eqref{ec:ad2} coincides with the classical $\delta N$ initial condition, given by $\bar{\phi}(N_i)+\delta\phi$, with $\delta\phi$ a Gaussian variable with variance \eqref{eq:varphi}. Choosing $N_i$ at a time such that the modes in the peak of $\mathcal{P}_\mathcal{\zeta}$ have frozen means that \eqref{ec:ad2} has already received its main contribution from the maximum of its integrand, and therefore the time dependence through the upper limit of integration is lost.\footnote{The lower limit of integration, which in practice corresponds to an infrared cutoff, is the same for the integral in \eqref{ec:ad2} and the one in \eqref{eq:varphi}.} From the point of view of classical $\delta N$, this scenario corresponds to a realization of the eternal CR case.\footnote{See the discussion in the last paragraph of Sec.\ \ref{sec:deltan}, where the dependence of the PDF of $\delta N$ on $N_i$ is lost (see Eq. \eqref{ec:pdeltaNfull}) provided it is enough for the peak modes of $\mathcal{P}_\mathcal{R}$ to have frozen.} From the point of view of stochastic $\delta N$, this corresponds to an approximate independence on the choice of $\sigma$ (Fig.~\ref{fig:stoch}, right panels), again provided it is chosen small enough for the modes of the peak to be coarse-grained once they have frozen.

If the attractor trajectory is not aligned with the CR line (as it happens for {\it Model 1}), the situation is different. While the noise vector kicks the field along the CR line, the drift vector tends to drive the system toward the attractor (where the grey lines converge). These two terms are not aligned and it is not possible to condense the system into a single variable. In general, the field and the momentum experience a random walk in the region of phase space between the CR line and the attractor trajectory which is not correctly captured by the eternal CR approximation (hence the change of slope of the exponential tail of the PDF of $\delta N$, see Fig.\ \ref{fig:stoch}, left panel). In principle, the equivalence between the classical and stochastic $\delta N$ formalisms is not exact, as the argument in \eqref{ec:ad0} does not extend to this case. Nevertheless, the fact that the largest stochastic kicks are clustered in a short time period (because of the power spectrum having a narrow peak), with a relatively constant $\eta$, still makes both approaches (classical and stochastic) very similar, provided $N_i$ (in classical $\delta N$) and $\sigma$ (in stochastic $\delta N$) are chosen consistently. Unlike in the previous case, there is a dependence of the PDF of $\delta N$ on $N_i$ for (classical $\delta N$) and $\sigma$ (for stochastic $\delta N$). The reason for this is that the earlier (or later) we chose $N_i$, the smaller (or larger) values of $\delta N$ will begin to probe the region of phase space in which the CR line and the attractor do not overlap anymore.

Let us indicate at this point that one might follow an alternative procedure to determine the direction (in phase space) of the kicks in the stochastic $\delta N$ formalism. Instead of evaluating \eqref{eq:noisevector} on the background trajectory at the time $N$, i.e.\ making the kick parallel to the CR line, one could evaluate it on the specific point in phase space to which the field has been stochastically driven at the time $N$, i.e.\
\begin{equation}
\begin{pmatrix}\xi_\phi\\\xi_\pi\end{pmatrix} \propto \hat \xi \left.\begin{pmatrix}1\\-\eta \end{pmatrix}\right|_{(\phi,\pi)} \,,
\label{eq:noisevector2}
\end{equation}
where $(\phi,\pi)$ denotes the (stochastic) field configuration at the time $N$, which in general will be outside of the unperturbed trajectory. Notice that, as long as that point in phase space corresponds to a negative $\eta$ (as it is the case, since this is the sign of $\eta$ on the phase space trajectories close to the attractor, see Fig.\ \ref{fig:phase}), $\mathcal{R}$ will remain frozen, and therefore it is consistent to use \eqref{eq:noiserelation} (with $\phi$ and $\pi$ evaluated on the corresponding point in phase space). This approach allows to partly account for the backreaction of the stochastic background on the noise, as the direction of the stochastic kicks depends on the point in phase space the field stochastically arrived to, and therefore it is path dependent. 
 
We will now argue that the results obtained by this method and the one described around \eqref{eq:noisevector} are almost identical. In order to obtain a quick estimate of the result obtained through this new procedure (avoiding stochastic calculations), we notice that, in practice, this procedure constrains the field to be stochastically kicked back and forth along the (possibly bent) attractor in phase space (see Fig.\ \ref{fig:phase}). Therefore, a reasonable approximation can be obtained by using the classical $\delta N$ formalism, but evaluating $\delta N(\delta \phi)$ along the bent attractor.\footnote{Notice the similarity with the unperturbed-trajectory approximation. However, unlike there, we are now constraining the field to move along a trajectory physically motivated by the freezing of $\mathcal{R}$.} The corresponding PDF is displayed in Fig.\ \ref{fig:stoch}, left panel. Notice the very little difference with respect to the classical and stochastic results obtained by kicking the field along the CR line with slope defined by $\eta$ evaluated on the background trajectory. The reason for such a small difference is that it is mainly the slope of the attractor that fixes the shape of the PDF. The effect of kicking the field away from the attractor is largely subdominant, precisely due to the attractor erasing it away. Moreover, in a model with a pure CR attractor (e.g.\ Fig.\ \ref{fig:phase}, right panel), both approaches are exactly equivalent, as it follows immediately from the coincidence of the attractor and the CR line in this case.

In the stochastic approach, the field and its momentum can, in principle, explore a region of parameter space inaccessible to the classical $\delta N$ formalism, namely, the red regions in Fig.~\ref{fig:phase}. This would be the case for stochastic realizations of the system in which the field receives kicks that take it to the region $\phi>\phi_{\max}$, but then it also receives enough kicks that bring it back to $\phi<\phi_{\max}$ and thus does not get stuck in the local minimum, as would happen classically. This is a difference between the two formalism (classical and stochastic). The stochastic $\delta N$ formalism could in principle describe these events more accurately. These effects are, however, largely mitigated in the case in which the power spectrum is very peaked, and thus the kicks occur in a narrow window of time. This is in fact the case for {\it Model 1}, for which both formalisms yield very close results, as we will now show.

We computed the PDF of $\delta N$ numerically by running $10^6$ realizations of the Langevin system in Eqs.\,(\ref{eq:stochasticinflation_first}, \ref{eq:stochasticinflation}). The results for {\it Model 1} are shown on the left panel of Fig.\,\ref{fig:stoch} for the two representative values $\sigma = 10^{-1}$ (orange crosses) and $10^{-2}$ (blue crosses). We have also solved the simplified equation (\ref{eq:stochastic1}) for $\sigma = 10^{-2}$ (red dots). The results for {\it Model 2} are shown on the right panel of the same figure. The black dashed line represents the eternal CR approximation of Eq.\,(\ref{eq:deltaphiCR}). The black solid line corresponds to the result obtained using the classical $\delta N$ formalism with momentum perturbations included, i.e.\,by following the procedure outlined below Eq.\,(\ref{eq:2shift}), with $N_i-N_{*}=4.2$.\footnote{The precise value of $N_i$ that makes the classical and stochastic calculations agree cannot be exactly determined a priori. However, as explained before, a reasonable estimate for it can be obtained by taking into account the following. For $\sigma=10^{-2}$, the mode corresponding to the peak of the power spectrum, $k_{\rm peak}$, is coarse-grained $-\log(10^{-2})\simeq  4.6$ $e$-folds after crossing the horizon, which, for our model, corresponds to an $N$ such that  $N-N_{*}=4.0$ $e$-folds. If the peak were infinitely narrow, we would therefore take $N_i-N_{*}=4.0$. However, since the peak of the spectrum has a certain width, we need $N_i-N_{*} \gtrsim 4.0$ to account for the freezing of modes close to the peak with $k\lesssim k_{\rm peak}$. Numerically, we found to best match to occur for $N_i-N_{*}=4.2$. \label{ft:1}  }

As shown in Fig.\,\ref{fig:stoch}, both the stochastic and classical versions of the $\delta N$ formalism (with momentum perturbations included) lead to extended, non-Gaussian tails for the PDF of the curvature perturbation. In the case of {\it Model 1}, the eternal CR approximation of Eq.\,(\ref{eq:deltaphiCR}) fails to correctly capture the tail behaviour due to the complexity of the phase space. {This is illustrated by the fact that the slope of the seemingly straight attractor trajectory close to the origin in the left panel of  Fig.\ \ref{fig:phase} differs from the slope of the late time attractor.} Similarly, the simplified, single-Langevin equation (\ref{eq:stochastic1}) deviates from the full result in the deep-tail regime for the reasons highlighted earlier.  On the other hand, for {\it Model 2} the phase space attractor coincides with the CR line, as shown in Fig.\,\ref{fig:phase}, and therefore all approaches yield essentially the same result.\footnote{We remark that, due to the steeper decrease of the PDF for this model, studying the deep-tail regime with the stochastic approach is much more costly, since probing large values of $\delta N$ requires many more realizations.} This figure also illustrates the importance of the choice of $\sigma$ in the stochastic approach, which mirrors the choice of $N_i$ in the standard $\delta N$ calculation. {\it Model 2} is less sensitive to this choice than {\it Model~1} due to the simpler structure of the phase space.

Before closing this section, let us comment on the effects of backreaction we mentioned earlier. As we stated before, a full treatment of stochastic inflation would involve recalculating the power spectrum, which appears in the amplitude of the noise coefficients, at every time step for each realization. As we already mentioned, this was done in \cite{Figueroa:2021zah}, and no deviation was found in the PDF with respect to the result obtained by computing the spectrum once by evaluating the coefficients in the background. The reason for this behaviour is simply that by the time the kicks begin to act on the field during the CR phase, the modes around the peak of the spectrum are coarse-grained and frozen. Thus, these modes of the power spectrum remain the same independently of changes in the background field and velocity. Moreover, the value at which these modes have frozen was determined by the background evolution in a period when the noise was negligible (i.e.\ before the peak of the power spectrum was coarse-grained). The only way this backreaction could have some effect is for very rare realizations in which the field is pushed back beyond the local maximum, and thus a second USR phase begins, momentarily changing the sign of $\eta$, until the field is driven back to CR so that inflation can end.\footnote{{We emphasize that with this statement we refer to realizations in which the field is pushed back beyond the local maximum and then driven back to the constant roll phase in such a way that inflation ends, so that they are included in the statistics. We do not refer to cases in which the inflaton is stuck in the local minimum, which are in fact not rare and make up $\mathcal{O}(1\%)$ of the total realizations. Once these realizations are removed from the statistics, the number of those falling into the former category is negligible.}} Even in this case, further super-Hubble evolution for $\zeta$ is highly unlikely due to the fact that, at this point, the second mode in Eq.\,(\ref{eq:second_mode}) has been decaying since the first USR phase ended, and thus the second USR phase would need to either occur very shortly after the first or have an extreme duration, and the probability of this happening is significantly  suppressed (as illustrated by our results). 

\section{Summary and conclusions} \label{sec:conc}

We have computed the PDF of the non-linear curvature perturbation on uniform-density hypersurfaces, $\zeta = \delta N$, in the presence of an USR phase using the classical and stochastic $\delta N$ formalisms. We have studied the degree of equivalence of these methods, as well as the validity of several approximations for both of them. We considered two different models, with two qualitatively different phase spaces, to understand how the morphology of phase space affects our conclusions.

In Sec.\ \ref{s:classical}, we applied the classical $\delta N$ formalism, imposing consistent conditions on field and momentum perturbations as per \eqref{eq:2shift}, to numerically compute the PDF of $\zeta$. As illustrated in Fig.\ \ref{fig:phase}, Eq.\ \eqref{eq:2shift} is equivalent to shifting the field at $N_i$ along a specific line in phase space that we refer to as the \emph{CR line}. We addressed the validity of two approximations:
\begin{itemize}
\item The \emph{eternal CR approximation}, which assumes that the field and momentum perturbations lie along the CR line and do not depart from it throughout their evolution. This approximation works well for models in which the CR line and the phase space attractor (approximately) overlap. In our examples, this is the case for \emph{Model 2}, but not for \emph{Model 1}. In the cases where this approximation holds, accurate analytical estimates of the PDF, as in Eq.\ \eqref{ec:pdeltaNfull}, can be obtained. 
\item The \emph{unperturbed trajectory approximation}, which assumes the perturbations are such that the field and its momentum evolve along the unperturbed trajectory in phase space. We found that such an approximation has a finite regime of validity, correspondent to the region where the CR line and the phase space attractor overlap, but is bound to eventually fail at some value of $\delta N$. Moreover, we argued that this failure manifests itself in the PDF of $\delta N$ in the form of a (spurious) bump.
\end{itemize}

In Sec.\ \ref{s:stoch}, we applied the stochastic $\delta N$ formalism, again imposing consistent conditions on field and momentum noises as per Eq.\ \eqref{eq:noiserelation}. In the same spirit of the unperturbed-trajectory approximation, we studied the outcome of simplifying the system of Langevin equations by making the field drift and diffuse along the unperturbed trajectory, as in \eqref{eq:stochastic1}. Similarly to the classical case, we found the approximation to fail eventually. Also in Sec.\ \ref{s:stoch}, we analysed the equivalence of the stochastic and classical $\delta N$ formalisms. From here, we extracted several important conclusions:
\begin{enumerate}
\item For the stochastic and classical $\delta N$ formalisms to be equivalent, a consistency relation needs to hold between $N_i$ (which indicates the initial hypersurface in the classical $\delta N$ formalism) and $\sigma$ (the coarse-graining parameter in the stochastic $\delta N$ formalism). For a peaked power spectrum, this relation is $k_{\rm peak}\approx \sigma a(N_i)H(N_i)$, where $k_{\rm peak}$ is the mode corresponding to the maximum of the power spectrum. The relation is physically meaningful, as both parameters indicate (in their respective formalisms) the delay from the horizon crossing of $k_{\rm peak}$ to its actual freezing. From a more technical point of view, this freezing is a necessary condition for the separate universe approach (underlying in both classical and stochastic $\delta N$ formalisms) to hold. The reason why the relation is approximate and not equal is that the peak of the spectrum is not strictly monochromatic, but slightly extended in $k$, as discussed in Footnote~\ref{ft:1}. 
\item For models with a phase space in which the attractor and the CR line approximately overlap (as in \emph{Model 2}), the equivalence between classical and stochastic $\delta N$ formalisms is exact, as the system is effectively described by dynamics in one dimension (given by the CR line). Moreover, in this case, both approaches are equivalent to the eternal CR approximation for classical $\delta N$, where the dependence on the final PDF of $\delta N$ (which has a single-slope exponential tail) on $N_i$ (equivalently, on $\sigma$) is lost, as it can be shown analytically (see \eqref{ec:pdeltaNfull}). This is in agreement with the results in \cite{Tomberg:2023kli}.
\item For models with a phase space in which the attractor has non-constant slope, and thus leaves the CR line, there is still an approximate equivalence between stochastic and classical $\delta N$ formalisms, ensured by the narrowness of the peak of the power spectrum (in short, this implies that most stochastic kicks occur during a short time range where $\eta$ of the attractor is almost constant). However, in this case, the tail of the PDF is given by a non-constant-slope exponential, related to the non-constant slope attractor in phase space. The position of the breaking points between different slopes depends on the choice of $N_i$ (or $\sigma$), therefore in this case the result is not $N_i-$independent. 
\end{enumerate}

A corollary of the last two points is that the non-Gaussian tails in the PDF of the curvature perturbation $\zeta$ can be reproduced without invoking the stochastic inflation formalism in models with a peaked power spectrum when momentum fluctuations are properly taken into account.  These tails are a consequence of the non-linear relation between $\zeta$ and $\delta\phi$. The reason why they also arise in stochastic $\delta N$ is because of the implementation of the  $\delta N$ formalism into stochastic inflation, not because of the stochastic formalism itself. From a practical point of view, this implies that the non-Gaussian tail of the PDF of $\zeta$ can be reproduced with a simple procedure that takes only a few lines of code and a negligible amount of computational power. Moreover, depending on the structure of the phase space, it may not be necessary to perform any numerical calculations at all, and the simple expression in Eq.\,(\ref{eq:deltaphiCR}) can be used to obtain the PDF analytically, as previously discussed in \cite{Tomberg:2023kli}.\footnote{Note that, in \cite{Tomberg:2023kli}, a deviation is found between the results obtained by using the simplified equation (\ref{eq:stochastic1}) and the eternal CR approximation (\ref{eq:deltaphiCR}). This difference disappears once one considers the full system (\ref{eq:stochasticinflation_first}, \ref{eq:stochasticinflation}) or, alternatively, the classical $\delta N$ formalism with momentum given by \eqref{eq:2shift}.}

As it has been emphasized, a key property to understand the relation between the classical and stochastic $\delta N$ formalisms and their possible simplifications is the constant slope (or lack thereof) of the phase space attractor for $\phi<\phi_{\max}$ (or equivalently, whether the attractor and the CR line lie on top of each other or not). As already mentioned, this property can be traced back to the second derivative of the potential. If it is close to constant around the local maximum, this maximum is well described by an inverted parabola, for which the equation of motion of the inflaton becomes linear (see \eqref{linear_eom}), with an attractor with (asymptotically) constant $\eta$. This is what happens in \emph{Model 2}. In \emph{Model 1}, the situation is the opposite. The second derivative of the potential varies around the local maximum, meaning that the equation of motion of the inflaton is non-linear, and the attractor is not characterized by a constant value of $\eta$. Likewise, the values of $\eta$ during USR and CR are not related by the Wands duality. In principle, it is to be expected that a single-field potential with a local minimum - local maximum configuration whose potential can be described with a small number of free parameters, like the radiative plateau potential in \cite{Ballesteros:2017fsr}, will be reasonably well approximated by an inverse parabola around its local maximum, and hence share the properties found for \emph{Model 2}. Reciprocally, potentials with many free parameters and complicated functional forms, or reverse-engineered from the evolution of some time-dependent quantity (as it is indeed the case of \emph{Model 1}), can develop the issues found in \emph{Model 1}. 

Another conclusion of our work is that, unless the unperturbed-trajectory approximation is imposed (which, as already discussed, is unreliable when probing the deep tail of the PDF of $\zeta$), the field might get stuck in the local minimum of the potential. This can happen equivalently in the classical $\delta N$ formalism ($\delta\phi$ and $\delta\pi$ being large enough for $(\bar{\phi}(N_i)+\delta\phi,\bar{\pi}(N_i)+\delta\pi)$ to be in the red region of Fig.\ \ref{fig:phase}) and the stochastic one (an accumulation of kicks parallel to the CR line bringing the field to the red region of Fig.\ \ref{fig:phase}, where the drift term carries it to the local minimum\footnote{In this case, there is a small probability of the noise bringing the field back again to the grey region of Fig.\ \ref{fig:phase}.}). Since the potential is designed to lead to a substantial slowdown of the scalar field, the probability of this happening is not, in principle, totally negligible. For the models under consideration, we found $\mathcal{O}(1\%)$ of the trajectories end up trapped in the minimum.

Finally, let us reiterate that the results obtained using both stochastic inflation and the classical $\delta N$ formalism rely crucially on the assumption that the field fluctuations $\delta\phi$ follow a Gaussian distribution. Studies about this assumption in the presence of a USR phase are scarce, particularly in the deep-tail regime, where these deviations would become more relevant. We remark, however, that the $\delta N$ formalism might provide an additional advantage in this context, since it is able to handle any non-Gaussian distribution for $\delta\phi$, as is clear from the simple relation in \eqref{changeprob}, whereas the Gaussianity of $\delta\phi$ is essentially built into the standard formulation of stochastic inflation in terms of Langevin equations and is therefore much more difficult to incorporate into the formalism.

\section*{Acknowledgments}
The work of GB and APR has been funded by the following grants: 1) Contrato de Atracci\'on de Talento (Modalidad 1) de la Comunidad de Madrid (Spain), 2017-T1/TIC-5520 and 2021-5A/TIC-20957, 2)  PID2021-124704NB-I00 funded by MCIN/AEI/10.13039/501100011033 and by ERDF A way of making Europe, 3) CNS2022-135613 MICIU/AEI/10.13039/501100011033 and by the European Union NextGenerationEU/PRTR, 4) the IFT Centro de Excelencia Severo
Ochoa Grant CEX2020-001007-S, funded by MCIN/AEI/10.13039/501100011033. APR has been supported by Universidad Aut\'onoma de Madrid with a PhD contract {\it contrato predoctoral para formaci\'on de personal investigador (FPI)}, call of 2021, and a grant \textit{Programa de ayudas UAM-Santander para la movilidad de j\'ovenes investigadores}, call of 2023. APR thanks DESY for hospitality during part of the realization of this work. MP and JR acknowledge support by the Deutsche Forschungsgemeinschaft (DFG, German Research Foundation) under Germany's Excellence Strategy – EXC 2121 “Quantum Universe” - 390833306. 
MP acknowledges support by the DFG Emmy Noether Grant No. PI 1933/1-1. We thank S.\ C\'espedes, J.\ Gamb\'in Egea and E. Tomberg for discussions.

\appendix

\section{The $\delta N$ formula}\label{ap:deltan}

In this appendix, we discuss in more detail the arguments and assumptions leading to the $\delta N$ formula \eqref{ec:formula_deltaN}. The starting point is the ADM decomposition of the metric, which we write following the standard parameterization used e.g.\ in \cite{Sugiyama:2012tj}
\begin{equation}\label{ec:ap1.1}
\diff s^2=-\alpha^2 \, \diff t^2 + \gamma_{ij}(\diff x^i+\beta^i \, \diff t)(\diff x^j+\beta^j \, \diff t)\,,
\end{equation}
where $\alpha$ and $\beta^i$ are respectively the lapse and shift functions, and
\begin{equation}\label{ec:ap1.2}
\gamma_{ij} = a^2\, e^{2\psi}\, \tilde{\gamma}_{ij}\,,
\quad 
\tilde{\gamma}_{ij} = (e^h)_{ij}, \quad \text{Tr}(h)=0\,.
\end{equation}
Given the normal 1-form to $t=$ constant hypersurfaces, $n_\mu=(-\alpha,0,0,0)$, and denoting $D_i$ the covariant derivative associated to the spatial metric $\gamma_{ij}$, the extrinsic curvature on spatial hypersurfaces can be written as \cite{Hamazaki:2008mh}
\begin{equation}\label{ec:ap1.3}
K_{ij}=\nabla_i n_j=\frac{1}{2\alpha} \left(-\dot{\gamma}_{ij} + D_i\beta_j + D_j\beta_i\right) = \frac{K}{3}\gamma_{ij} + a^2e^{2\psi}\tilde{A}_{ij}\,,
\end{equation}
where $\tilde{A}_{ij}$ is traceless and $K$ is the trace of the extrinsic curvature, given by
\begin{equation}\label{ec:ap1.4}
K=-\frac{1}{\alpha}\left[3(H+\dot{\psi})-D_i\beta^i\right], \quad H(t)=\dot{a}
(t)/a(t)\,.
\end{equation}
The trace of the extrinsic curvature can be used to define the \emph{generalized number of $e$-folds}, 
\begin{equation}\label{ec:ap1.5}
\diff \mathcal{N} = -\frac{\alpha \,K}{3}\diff t\,,
\end{equation}
which reduces to $\diff N=H\,\diff t$ for a FLRW metric.

In the ADM formalism, the quantities $\psi$, $\tilde{A}_{ij}$, $K$ and $\tilde{\gamma}_{ij}$, together with the inflaton $\phi$, are dynamical degrees of freedom. Their equations of motion are supplemented with two constraint equations obtained varying the action \eqref{inflaset} with respect to the lapse and the shift \cite{Salopek:1990jq}, known as \emph{Hamiltonian} and \emph{momentum constraints} \cite{Hamazaki:2008mh, Sugiyama:2012tj}. Instead of expanding the equations in powers of the $\psi$, $\tilde{A}_{ij}$, $K$, $\tilde{\gamma}_{ij}$ and $\delta\phi=\phi-\bar{\phi}(t)$ \footnote{In this appendix, $\bar{\phi}$ denotes the solution of the scalar field equation when \eqref{ec:ap1.1} is a FLRW metric.} --which would then be assumed to be small fluctuations, as it is done in standard perturbation theory--, we can expand them in spatial gradients, denoting with powers of $\epsilon_g$ the different orders in the gradient expansion. Working in a gauge in which $h_{ij}$ is transverse and assuming that the anisotropic stress is $\mathcal{O}(\epsilon_g^2)$, one can show \cite{Garriga:2015tea, Sugiyama:2012tj} that $\tilde{A}_{ij}$, $\dot{h}_{ij}$ and $\partial_j\beta^i$ are also $\mathcal{O}(\epsilon_g^2)$. Using this result, and changing the time variable to the generalized number of $e$-folds, one finds that the Hamiltonian constraint and the equation of motion for $\phi$ have the same functional forms as their unperturbed counterparts (for which we assume a FLRW metric) at lowest order in the gradient expansion:\footnote{Second order gradients are \emph{not} negligible in an USR phase, already at linear order in perturbations \cite{Romano:2015vxz, Cruces:2022dom}. This is consistent with the fact that both $N_i$ (in the classical $\delta N$ formalism) and $\sigma$ (in the stochastic one) have to be chosen such that modes exiting the horizon during USR do not affect the unperturbed solutions (via classical perturbations or stochastic kicks) until they freeze after the USR phase.} 
\begin{align}
3\tilde{H}^2 -\frac{\tilde{H}^2(\partial_{\mathcal{N}}\phi)^2}{2}- V & = \mathcal{O}(\epsilon_g^2)\,,\label{ec:ap1.6.1}\\
\tilde{H}\,\partial_{\mathcal{N}}(\tilde{H}\,\partial_{\mathcal{N}}\phi)+3\tilde{H}^2\partial_{\mathcal{N}}\phi+\partial_\phi V & =\mathcal{O}(\epsilon_g^2)\,,\label{ec:ap1.6.2}
\end{align}
where $\tilde{H}=-K/3$. 
This non-linear scheme to treat non-linear cosmological perturbations of large wavelength is referred to as the \emph{separate universe approach}.

Integrating \eqref{ec:ap1.5} and neglecting $\mathcal{O}(\epsilon_g^2)$ terms, we get \cite{Lyth:2004gb} \footnote{Aside from being mathematically convenient for the argument, the integral of \eqref{ec:ap1.5} can be physically interpreted (up to $\mathcal{O}(\epsilon_g^2)$ corrections) as the expansion along a comoving congruence of geodesics \cite{Lyth:2004gb, Garriga:2015tea}.} 
\begin{equation}\label{ec:ap1.7}
\psi(t_2,\bm{x})- \psi(t_1, \bm{x}) = \mathcal{N}(t_1\to t_2,\bm{x}) - N(t_1 \to t_2) + \mathcal{O}(\epsilon_g^2)\,,  
\end{equation}
where
\begin{align}
\mathcal{N}(t_1\to t_2,\bm{x})=\int_{t_1}^{t_2}\diff \mathcal{N}\,,
\end{align}
and $N(t_1\to t_2)=\int_{t_1}^{t_2}\diff t\, H$ corresponds to the number of $e$-folds among the $t=t_1$ and $t=t_2$ hypersurfaces in the unperturbed (FLRW) spacetime. No specific gauge has been introduced so far. We now choose a gauge that interpolates between the \emph{flat gauge} ($\partial^i h_{ij}=\psi=0$) hypersurface $\Sigma_{\text{f},t_1}$ at $t=t_1$ and the \emph{uniform density gauge} ($\partial^i h_{ij}=\delta\rho=0$) hypersurface $\Sigma_{\text{ud},t_2}$ at $t=t_2$. Applying this gauge choice in \eqref{ec:ap1.7}, we get
\begin{equation}\label{ec:ap1.8}
\left.\psi(t_2, \bm{x})\right|_{\partial^ih_{ij}=\delta\rho=0} = \mathcal{N}(\Sigma_{\text{f},t_1} \to \Sigma_{\text{ud},t_2}, \bm{x})-N(t_1\to t_2)+\mathcal{O}(\epsilon_g^2)\,.
\end{equation}
The left-hand side is, by definition, the \emph{non-linear perturbation}\footnote{The non-linear $\zeta$ is usually defined in this way \cite{Sugiyama:2012tj, Garriga:2015tea}. See, however, the discussion in \cite{Cruces:2022dom}. An alternative definition is presented in \cite{Lyth:2004gb}, which is however \emph{not} gauge invariant and only reduces to the standard definition in the perfectly adiabatic limit $p=p(\rho)$. } $\zeta(t_2,x^i)$. Hence,
\begin{equation}\label{ec:ap1.9}
\zeta(t_2; x^i) = \mathcal{N}(\Sigma_{\text{f},t_1} \to \Sigma_{\text{ud},t_2}, \bm{x})-N(t_1\to t_2) + \mathcal{O}(\epsilon_g^2)\,.
\end{equation} 
In order for \eqref{ec:ap1.9} to be of practical use, we need a method to compute $\mathcal{N}(\Sigma_{\text{f},t_1} \to \Sigma_{\text{ud},t_2}; \bm{x})$, which is provided by the separate universe approach.\footnote{Notice that the separate universe approach is applicable to \eqref{ec:ap1.9} because both the flat and uniform-density gauges imply $\partial^i h_{ij}=0$, which is necessary to prove the separate universe approach, see \cite{Sugiyama:2012tj}.} Thanks to the explicit analogy between eqs.\ \eq{ec:ap1.6.1} and \eq{ec:ap1.6.2} and the ones for the homogeneous $\phi$, the dependence of the non-linear $\phi(\mathcal{N})$ and $\pi(\mathcal{N})$ on $\mathcal{N}$ is the same as the unperturbed $\bar{\phi}(N)$ and $\bar{\pi}(N)$ on $N$. This implies that the functional form of $\mathcal{N}(\phi,\pi)$ is the same as that of $N(\bar{\phi},\bar{\pi})$.  As a consequence, 
$\mathcal{N}(\phi,\pi)$ and $N(\bar{\phi},\bar{\pi})$ may differ \emph{only} due to their respective boundary conditions. 

In Eq.\ \eqref{ec:ap1.9}, the boundary conditions are specified by the gauge choice at $t=t_1$ and $t=t_2$. In particular, at $t=t_1$, we impose the flat gauge, therefore $\phi(t_1, \bm{x})=\bar{\phi}(t_1)+\delta\phi_{\text{f}}(t_1,\bm{x})$, $\pi(t_1, \bm{x})=\bar{\pi}(t_1)+\delta\pi_{\text{f}}(t_1;\bm{x})$ where $\delta\phi_{\text{f}}$ and $\delta\pi_{\text{f}}$ are the field and velocity perturbations computed in the flat gauge. In practice, $\delta\phi_{\text{f}}$ and $\delta\pi_{\text{f}}$ are computed (approximately) by using linear perturbation theory in the flat gauge. At $t=t_2$, we impose the uniform density gauge, therefore $\delta\rho(t_2, \bm{x})=0$ for all $\bm{x}$. 

As argued in Sec.\ \ref{sec:deltan}, momentum perturbations in the initial hypersurface are not negligible in general. Regarding the final hypersurface, the velocity perturbations can be safely neglected.\footnote{See \cite{Pi:2022ysn} for a full calculation involving field and momentum perturbations at both initial and final hypersurfaces.} This is because:
\begin{enumerate}
\item The field has evolved long enough in the presence of a CR attractor for it to effectively reach the unperturbed trajectory, i.e.\ the evolution of the field dynamically erases $\delta\pi_{\text{ud}}(t_2,\bm{x})$.
\item The dependence on $\delta\pi_{\text{ud}}(t_2,\bm{x})$ is through $\delta\rho[\delta\phi_{\text{ud}}(t_2,\bm{x}),\delta\pi_{\text{ud}}(t_2,\bm{x})]=0$. Along the unperturbed trajectory, this is suppressed by the slow-rolling of the inflaton (the main contribution to the energy density is the potential, i.e. the field value $\delta\phi_{\text{ud}}(t_2,\bm{x})$). 
\end{enumerate}
This implies that the gauge condition in the last hypersurface (on top of $\partial^i h_{ij}$) is
\begin{equation}\label{ec:3.51}
\delta\rho[\delta\phi_{\text{ud}}(t_2,\bm{x}),\delta\pi_{\text{ud}}(t_2,\bm{x})]\simeq \delta\rho[\delta\phi_{\text{ud}}(t_2,\bm{x})]=0\,.
\end{equation}
Neglecting gradient-suppressed terms, $\delta\rho=0 \iff \delta\phi=0$.\footnote{
For $\delta\phi=0$, the background energy density is 
$\rho = \dot{\bar{\phi}}^2/(2\alpha^2)+V$.
Writing $\alpha = 1+\delta\alpha$, 
then, the energy density fluctuation in the gauge in which $\delta\phi=0$ is $\left.\delta\rho\right|_{\delta\phi=0}=-\delta\alpha\,\dot{\bar	{\phi}}^2(1+\mathcal{O}(\delta\alpha))$. Since $\delta\alpha$ is gradient suppressed in the gauge where $\delta\phi=0$ \cite{Maldacena:2002vr}, we conclude that $\delta\rho=0\iff\delta\phi=0$ provided gradients are negligible.
}
Therefore, since we assume the gradient expansion to hold at $t_2$,
\begin{equation}
\delta\phi_{\text{ud}}(t_2,\bm{x})=0\Rightarrow \phi(t_2,\bm{x})=\bar{\phi}(t_2)\,,
\end{equation}
i.e.\ the second hypersurface can be taken as an \emph{uniform-field slice}, which is convenient from a practical point of view.

Taking the above into account, the separate universe approach provides the following prescription to compute $\mathcal{N}(\Sigma_{{\text{f}},t_1} \to \Sigma_{\text{ud},t_2},\bm{x})$: solve the \emph{unperturbed} equations of motion with initial conditions $\bar{\phi}(t_1)+\delta\phi_{\text{f}}$, $\bar{\pi}(t_1)+\delta\pi_{\text{f}}$, and then compute the number of $e$-folds until $\phi=\bar{\phi}(t_2)$ along said solution. The subtrahend in  \eqref{ec:ap1.9} is similarly computed, but using $\bar{\phi}(t_1)$ and $\bar{\pi}(t_1)$ as initial conditions. In summary,
\begin{align}\label{ec:master_deltaN} \nonumber
\zeta(t_2,\bm{x})  =\, & N\left[\{\bar{\phi}(t_1)+\delta\phi_{\text{f}}(t_1,\bm{x}),\bar{\pi}(t_1)+\delta\pi_{\text{f}}(t_1,\bm{x})\}\to \bar{\phi}(t_2)\right]\\
 & -N\left[\{\bar{\phi}(t_1), \bar{\pi}(t_1)\}\to \bar{\phi}(t_2) \right]\,,
\end{align}
where $N$ denotes the functional form of $e$-folds as a function of field value and velocity derived from the unperturbed equations of motion. \



\bibliographystyle{utphys}
\bibliography{references-stochastic-delta-N} 

\end{document}